\documentclass[frontmatterverbose,
reprint, 
showkeys, nofootinbib, amsmath,amssymb, 
superscriptaddress, aps, prb, floatfix]{revtex4-2}
\usepackage[T1]{fontenc}
\usepackage{newtxtext, newtxmath}
\usepackage{orcidlink}
\usepackage{graphicx}
\usepackage{bm}
\usepackage{bbm}
\usepackage{dsfont}
\usepackage{physics}
\usepackage[byauthor, separator = {;\ }]{credits}
\usepackage{enumitem}
\usepackage{ulem, soul, tcolorbox}
\tcbuselibrary{most}

\begin{document}
\title{Integrable Floquet Time Crystals in One Dimension}
\author{Rahul Chandra\,\orcidlink{0000-0001-6328-2461}}
\credit{RC}{Conceptualization, Data curation, Project administration, Software, Visualization, Writing -- original draft}
\affiliation{Department of Physics, The University of Burdwan, Barddhaman, West Bengal 713104, India}

\author{Mahbub Rahaman\,\orcidlink{0000-0002-7455-2364}}
\credit{MR}{Investigation, Resources, Validation, Writing -- review \& editing}
\affiliation{Harish-Chandra Research Institute, A CI of Homi Bhabha National Institute, Prayagraj, Uttar Pradesh 211019, India}

\author{Soumyabroto Majumder\,\orcidlink{0009-0005-6929-6414}}
\credit{SM}{Resources, Validation, Visualization}
\affiliation{Department of Physics, The University of Burdwan, Barddhaman, West Bengal 713104, India}

\author{Analabha Roy\,\orcidlink{0000-0002-4797-0624}}
\email[Corresponding author: Email ]{daneel@utexas.edu}
\credit{AR}{Conceptualization, Data curation, Project administration, Formal analysis, Investigation, Methodology, Resources, Software, Supervision, Visualization, Writing -- original draft}
\affiliation{Department of Physics, The University of Burdwan, Barddhaman, West Bengal 713104, India}

\author{Sujit Sarkar}
\credit{SS}{Conceptualization, Formal analysis, Funding acquisition, Investigation, Methodology, Supervision, Validation, Writing -- review \& editing}
\affiliation{Theoretical Sciences Department, Poornaprajna Institute of Scientific Research, Bangalore 562164, India}
\date{\today}
\begin{abstract}
We demonstrate the realization of a Discrete Time-Crystal (DTC) phase in a family of periodically driven, one-dimensional quadratic lattice Hamiltonians that can be obtained using spin chains. These interactions preserve integrability while opening controllable gaps at resonant quasienergies and pinning the emergent quasienergy modes that are responsible for subharmonics. We demonstrate that the DTC phase is rigid in the parameter space of transverse field and an additional interaction like Next-Nearest-Neighbor (NNN) coupling strength, with the drive frequency optimized to produce the strongest subharmonic response. We also provide a detailed phase diagram of the model, exhibiting a Floquet Paramagnet (FPM) phase, as well as sharp quantum phase transitions between the FPM and the DTC. Finite-size scaling of the Floquet quasienergy splitting between the emergent subharmonic mode and its conjugate shows that the DTC lifetime diverges exponentially with system size. Thus, our work establishes a novel mechanism for achieving robust long-lived DTCs in one dimension. Motivation for this work stems from the limitations of disorder-based stabilization schemes that rely on many-body localization and exhibit only prethermal or finite-lived plateaus, eventually restoring ergodicity. Disorder-free routes are therefore highly desirable. Integrable (or Floquet-integrable) systems provide an attractive alternative because their extensive set of conserved quantities and constrained scattering strongly restrict thermalization channels. Our construction exploits these integrable restrictions together with longer-range NNN engineering to produce a clean, robust DTC that avoids the prethermal fragility of disordered realizations.
\end{abstract}
\keywords{Discrete Time-Crystal, Floquet Time Crystal, Integrable Systems, Spin Chains, Quantum Phase Transitions}
\maketitle
\section{\label{sec:intro}Introduction}
Spontaneous symmetry breaking (SSB) is the key principle that governs emergent order in many-body systems. Following Wilczek's proposal of \emph{time-crystals}~\cite{tc:wilczek:2012,tc:shapere:2012}, which envisioned phases breaking the continuous-time translation symmetry in equilibrium, a sequence of no-go theorems established that such equilibrium-time crystals cannot arise in generic short-range Hamiltonians~\cite{tc:patrick:comment:2013,tc:watanabe:oshikawa:2015}, steering interest toward intrinsically non-equilibrium settings. In periodically driven (Floquet) systems, for instance, the \textit{Discrete} time-translation symmetry may break spontaneously, producing Discrete Time-Crystals (DTC) with robust subharmonic response, spatiotemporal long-range order, and rigidity to perturbations~\cite{else:bauer:naik:2016}. These criteria, now standard in the community, underlie both theory~\cite{tc:rmp:colloquium:2023,else:2020} and experiment~\cite{choi:2017,zhang:2017} and provide the baseline against which we position our results.

The central challenge underlying stable subharmonic (period-doubled) response in Discrete Time-Crystals is preventing \emph{melting} of the subharmonic plateau under generic perturbations and finite-size or finite-time effects~\cite{kyprianidis:2021,pizzi:2020,liu:dtc:2023}. Early realizations of driven Ising-like chains employed disorder-induced many-body localization (MBL)~\cite{nandkishore:2015,zhang:2017,randall:2021} to inhibit heating between nominal $\pi$-pulses in spin-echo-style Floquet sequences~\cite{li:2007,else:bauer:naik:2016,yao:potter:potirniche:vishwanath:2017}, thus stabilizing emergent $2T$ oscillations. However, such disorder-stabilized DTCs typically enter a \emph{prethermal} regime~\cite{kyprianidis:2021,stasiuk:2023,vu:2023} with an extended but ultimately finite-lived subharmonic plateau that decays once residual interactions or processes in rare regions restore ergodicity~\cite{krajewski:2022,ha:2023}. (See, e.g., prethermal plateau analyses and lifetime scaling discussions in works associated with Huse and collaborators.~\cite{birnkammer:2022,luitz:2020,prethermal:pizzi:2021,machado:2020, nandkishore:2015})

These limitations motivate the search for \emph{disorder-free, clean, and symmetry-protected} routes to persistent subharmonics~\cite{Santini2022,pizzi:2021,Li:2025}. Continuous (as opposed to discrete) time crystal behavior in driven or pumped Bose–Einstein condensates (e.g. Ke{\ss}ler, Hemmerich and related cavity/BEC platforms)~\cite{kebler:2019,huetado:2020, kebler:2021,dtcbec:wang:2021,wang:2021,Bull2022}exemplifies an alternative mechanism. Other proposals involve quantum many-body scars~\cite{maskara:2021,deng:2023,chandran:2023,pizzi:2025}, dynamical many-body
localization~\cite{Rahaman:2024}, flat-bands~\cite{rahaman:2026} and strong Stark potentials~\cite{liu:dtc:2023,kshetrimayum:2020}. In parallel, a distinct disorder-free pathway~\cite{banerjee:2023,Haldar2022} emerges in \emph{integrable or Floquet-integrable} systems~\cite{integrable:floquet:gritsev:2017,banerjee:2023,qtckt:vernier:2024}: exact (or quasi-exact) conservation laws\cite{znidaric:2025,muller:2023}, strong (or almost strong) $\pi$ modes~\cite{shtanko:2020,floquet:yates:2021}, and constrained quasiparticle scattering~\cite{denardis:2019,giergiel:2018} combine to pin subharmonic responses without the need for localization~\cite{li:2007,anisur:2025,axs:2024}. Recent work~\cite{floquet:yates:2021,yates:2022,qtckt:vernier:2024} shows that one-dimensional integrable systems can be realized in quantum simulators~\cite{georgescu:2014,blatt:2012,frey:2022,Lu:2025, Weaving:2025}, and quantum dynamics can be engineered by Trotterization~\cite{trotter:pastori:2022,trotter:zhao:2023}.

Earlier demonstrations of disorder-free time-crystalline behavior in integrable models include the driven Lipkin–Meshkov–Glick (LMG) system~\cite{Russomanno:2017}, in which period-doubling originates from symmetry-broken collective states and associated quasi-energy doublets; periodically driven BCS condensates that exhibit parametric resonances~\cite{OjedaCollado:2021} and realize commensurate time-crystal phases within Arnold tongues; and $p-$spin models with infinite-range interactions that support higher-order subharmonic responses~\cite{MuozArias:2022}. These realizations, however, are subject to important limitations. In the driven LMG model, the periodic drive destroys integrability~\cite{rahamanlmg:2024}, while the other setups rely on semiclassical limits. In contrast, the present work investigates the exact quantum dynamics of a fully integrable one-dimensional lattice model, wherein time-crystalline behavior arises from momentum-resolved quasiparticle dynamics and Floquet band engineering enabled by next-nearest-neighbour couplings.

Our earlier work~\cite{inttc:chandra:2024} demonstrated the onset of Discrete Time-Crystalline order in \emph{higher-dimensional} integrable lattices, but revealed a fragility upon dimensional reduction to strictly one dimension. The key bottleneck was an insufficient parameter manifold to simultaneously (i) satisfy resonant mode pinning conditions, (ii) open and control quasienergy gaps at the relevant $\pi$ (or near-$\pi$) quasienergies, and (iii) suppress dephasing channels associated with nearby continua in momentum space. In the present work, we remedy that dimensional fragility by introducing controlled next-nearest-neighbor (NNN) (and effectively longer-range multi-spin) couplings that (a) preserve integrability, (b) enlarge the tunable parameter space, and (c) generate adjustable Bogoliubov (and Floquet) gap structures that robustly isolate a subharmonic mode. This converts a previously \emph{fragile} (fine-tuned) subharmonic into a \emph{rigid} phase protected by symmetry and sustained over an extended region in the parameter space.

Conceptually, the approach illustrates that \emph{NNN} extensions, rather than truly long-range algebraic tails, suffice for stabilization in a clean, translationally invariant setting. The added NNN (or three-spin Jordan–Wigner–generated) structures supply just enough dispersion engineering to ensure (i) controllable quasienergy gap opening at resonance, (ii) momentum-selective mode pinning (a $\pi$-mode analog) and (iii) suppression of multimode dephasing by reducing accidental degeneracies. Thus, integrability is leveraged not only as an analytical simplifyer but also as an active stabilizing mechanism for Floquet subharmonics, offering a disorder-free counterpart to MBL- or prethermality-based stabilization routes.
We therefore position this work at the intersection of: (1) the ongoing program of disorder-free or clean DTC realization, (2) the exploitation of strong / almost strong mode physics in integrable and Floquet-integrable spin and fermionic chains, and (3) dimensional reduction strategies that retain controllable resilience of subharmonic order without invoking long-range interactions or open-system engineering. The resulting phase diagram, which contains robust DTC and Floquet Paramagnet (FPM) regimes, provides a unified framework for testing how integrability shapes temporal symmetry breaking, quasienergy topology, and finite-size scaling of melting times.

Concretely, in this manuscript we map the Floquet phase diagram in a manifold of Hamiltonian parameters consisting of the periodic drive amplitude $g_0$ and, for concreteness, an NNN coupling strength $\lambda$ added to the Ising spin chain. We optimize the drive frequency to maximize the stability of the DTC phase and identify a contiguous DTC region bounded by analytically obtained Floquet gapless points at both high-symmetry and non-high-symmetry momenta; adjacent to these boundaries we find FPM behavior. Thus, a rich phase diagram emerges with sharp quantum transitions involving DTC in an exact, integrable, closed quantum system that goes beyond the simple symmetry-broken/thermal phase transitions seen in MBL systems\cite{tc:rmp:colloquium:2023}, and in analogy with recently studied phase transitions in fractal time crystals in the mean-field limit\cite{giachetti:2023}, as well as open quantum systems\cite{mattes:2023}.

Dynamically, the DTC is diagnosed by near-unity long-time-averaged fidelity at optimal momentum $k_0$ and a robust stroboscopic correlation $\overline{C_z}\sim\mathcal{O}(1)$, while the FPM shows suppressed fidelity and qualitatively different temporal signatures.
Further analysis of these phases involves scaling hypothesis tests on temporal long-range correlations, which demonstrate the onset of \emph{off-diagonal long-range order} (ODLRO) in time for the DTC phase and rapid decay of these correlations near criticality in the FPM phase, justifying the nomenclature chosen. Finally, we extract a splitting $\delta\Omega(N)$ of the subharmonic peak from finite-size analyzes of the stroboscopic signal (FFT of $C_z(k_0^R,nT)$), and fit scaling laws $\delta\Omega\sim N^{\alpha}$: deep in the DTC we obtain algebraic scaling with $\alpha\approx-1$ (beat period $\sim N$), while in FPM the splitting either broadens, plateaus, or yields unstable exponents. These scaling results are obtained robustly through two Lorentzian peak fits and regression by Random Sampled Consensus (RANSAC). RANSAC is a robust algorithm used in machine learning~\cite{fischler:ransac,scikit-learn} that suppresses outliers from failed double-peak fits or resolution-limited spectra. Successful hypothesis tests have provided very clear finite-size distinctions between the truly rigid DTC order and FPM behavior.

Beyond establishing stability, our analysis clarifies how integrability is responsible for the rigidity of the DTC phase. In particular, we connect subharmonic oscillations to (a) the structure of conserved charges in the quadratic fermion description, (b) $\pi$-mode pinning akin to strong-mode physics in Floquet-integrable chains and circuits, and (c) finite-size dependence of melting of the DTC at accessible drive frequencies that complement, rather than substitute for, exact integrable control. This synthesis allows for the realization of DTCs by methods previously developed for one-dimensional integrable chains and circuits, providing a unified route to subharmonic order in clean systems.

Thus, our results should be read alongside three intertwined lines of inquiry.
\begin{enumerate}
    \item Clean DTCs without disorder: high-frequency prethermal DTCs in generic systems and domain-wall-confined DTCs in kicked chains supply complementary stabilization routes; our models realize rigidity without relying on either MBL or all-to-all interactions.
    \item Integrable / Floquet-integrable diagnostics: Strong and almost strong mode analyses in Floquet spin chains and integrable XXZ circuits motivate our $\pi$-mode pinning picture and guide our spectral tests. 
    \item Higher-dimensional integrable free fermions: \cite{inttc:chandra:2024} established the feasibility of higher-D integrable DTCs; we extend that framework by showing that NNN couplings within the integrable class enhance stability and broaden parameter windows for subharmonic order.
\end{enumerate}

Our work is organized as follows. In Section~\ref{sec:isingchain}, we introduce the basic model and its dynamics. In Section~\ref{sec:rigidity}, we present analytical and numerical explorations of the dynamics of the order parameter, building a complete phase profile of the system, and demonstrating quantum phase transitions between the DTC and FPM phases. We also elucidate how integrability ensures rigidity of the DTC phase in a three-parameter space of drive frequency, transverse field, and NNN coupling strength. In Section~\ref{sec:finitesize}, we present the finite-size scaling of stroboscopic correlations, compared against exact numerics. Finally, we present our conclusions and outlook.

\section{The Basic Model and Dynamics}
\label{sec:isingchain}
Consider the class of spinless free-fermionic models represented by a quadratic Bogoliubov-de Gennes (BdG) Hamiltonian summed over momentum pairs in the First Brillouin Zone (FBZ) of a one-dimensional lattice:
\begin{equation}
    H = \sum_{k,-k} \hat{\Psi}^{\dagger}_{k} H^{\;}_{k}(t) \hat{\Psi}^{\;}_{k}
    \label{eq:quadratic}
\end{equation}
where, \( \hat{\Psi}^{\dagger}_{k} = \pmqty{c^{\dagger}_{k} & c^{\;}_{-k} } \) is a Nambu spinor and \(c^{\dagger}_{k}\; (c^{\;}_{k})\) are fermionic creation (annihilation) operators. The BdG Hamiltonian $H_{k}(t)$ consists of a time-dependent $2\times 2$ traceless Hermitian matrix given by $\bm{r}_{k}(t)\cdot\bm{\tau}$, where the vector of Pauli matrices $\bm{\tau} = \tau_{1}\hat{x} + \tau_2\hat{y} + \tau_3\hat{z}$ consists of three matrices $\tau_{1,2,3}$ given by
$\tau_{1} = \smqty(\pmat{1}),\tau_2 =\smqty(\pmat{2}), \tau_3 = \smqty(\pmat{3})$, and the Bloch vector $\bm{r}_{k}(t)\sim \Xi_{k}(t) \hat{x} + \Upsilon_{k}(t) \hat{z}$. Here, $\Xi_{k}(t), \Upsilon_{k}(t)$ are real-time-dependent functions of momenta and are described by system-specific Hamiltonian parameters. Substituting into~\eqref{eq:quadratic}, the full Hamiltonian is expanded as:
\begin{multline}
    H(t) = \sum_{k,-k} \bigg[ \Upsilon_{k}(t) \left(c^{\dagger}_{k} c^{\;}_{k} + c^\dagger_{-k}c^{\;}_{-k}\right) + \\
    \Xi_{k}(t) \left(c^\dagger_{k} c^\dagger_{-k} + c^{\;}_{-k}c^{\;}_{k} \right)\bigg].
    \label{eq:Hamiltonian}
\end{multline}

The Hamiltonian in \ref{eq:Hamiltonian} can be rewritten in terms of Bogoliubov quasiparticles~\cite{bogie:1958,valatin:1958,zhu:2016} (Bogolons) that are annihilated by fermionic operators $\gamma_{k}$, where
\begin{align}
\gamma^{\;}_{k} &= c^{\;}_{k} \cos{\varkappa_{k}}  + c^\dagger_{-k} \sin{\varkappa_{k}}, \nonumber \\
\gamma^{\;}_{-k} &= c^{\;}_{-k} \cos{\varkappa_{k}}  - c^\dagger_{k} \sin{\varkappa_{k}},
\label{eq:bogolons}
\end{align}
with $ \tan{(2\varkappa_{k})} = \Xi_{k}/\Upsilon_{k}$. Substituting into Eqs.~\eqref{eq:bogolons} and~\eqref{eq:Hamiltonian} yields
\begin{equation}
H = \sum_{k,-k} \bar{E_{k}} \left(  \gamma^{\dagger}_{k} \gamma^{\;}_{k} + \gamma^{\dagger}_{-k} \gamma^{\;}_{-k} -1 \right)
\label{eq:BogHamiltonian}
\end{equation}
with energy eigenvalues $\bar{E_{k}} = \sqrt{\Upsilon^{2}_{k} + \Xi^{2}_{k}}$. Thus, the Hamiltonian conserves the Bogolon number at each momentum, given by the observable $\gamma^{\dagger}_{k} \gamma^{\;}_{k}$ : \( \comm{H}{\gamma^{\dagger}_{k} \gamma^{\;}_{k}} = 0\). This yields an extensive set of independent conserved quantities $\{\gamma^{\dagger}_{k} \gamma^{\;}_{k}\; \forall k\in\text{FBZ}\}$ for half of the FBZ, one for each positive $k$. Hence, the Hamiltonian is integrable, as it can be described as an ideal gas of Bogolons that scatters without diffraction~\cite{sutherland:beautiful-models}.

Now, suppose that the time dependencies are chosen in such a way that the matrix $H_{k}$ alternates between two matrices $\left|H_{1}\right|_{k}$ and $\left|H_{2}\right|_{k}$ with time period $T$ (frequency $\omega = 2\pi/T$) and $50\%$ duty cycle, where
\begin{align}
    \left|H_{1}\right|_{k} &=\left(g_{0}-b_{k}\right)\left(c^{\dagger}_{k} c^{\;}_{k}+c^{\dagger}_{-k}c^{\;}_{-k}\right)+\Delta_{k}\left(c^{\dagger}_{k} c^{\dagger}_{-k}+\text { h.c. }\right) \nonumber\\
    \left|H_2\right|_{k} &=g_{1}\left(c^{\dagger}_{k} c^{\;}_{k}+c^{\dagger}_{-k} c^{\;}_{-k}\right).
\label{eq:H12k}
\end{align}
The Hamiltonian consists of two parts: (1) free-fermion terms $c^{\dagger}_{k} c^{\;}_{k}$ that provide kinetic energies $g_{0}-b_{k}$ for half the period, producing a group velocity dispersion $b^\prime_{k}$ and a flat-band dispersion without group velocity $g_{1}$ for the other half; (2) strongly correlated Cooper pair terms $c^\dagger_{k} c^\dagger_{-k}$ with their Hermitian conjugates, involving interaction energy $\Delta_{k}$ for half the period. If the symmetry-breaking field amplitude $g_{0}$ ensures a momentum $k_{0}$ where $g_{0}=b_{k_{0}}$ and $\omega$ gives enough time for $\left|H_{1}\right|_{k_{0}}$ to generate a Cooper pair state from vacuum at the momentum $\pm k_{0}$ given by $\ket{k_{0}, -k_{0}} = c^\dagger_{k_{0}}c^\dagger_{-k_{0}}\ket{0}$, two criteria must be met. 
\begin{equation}
    g_{0} =b_{k_{0}}, \qq{and}
    \omega = 2\Delta_{k_{0}}.
\label{eq:tc:criteria}
\end{equation}
At time $t=T/2$ immediately following the first duty cycle, a Cooper pair with momenta $\pm k_{0}$ materializes from vacuum due to the action of the Hamiltonian $\left|H_{1}\right|_{k_{0}}$. This state then experiences ballistic evolution through the second duty cycle under $\left|H_{2}\right|_{k_{0}}$ until $t=T$. In the subsequent period, the pair is annihilated by $\left|H_{1}\right|_{k_{0}}$ at $t=3T/2$, initiating another ballistic transition of the resulting vacuum with $\left|H_{2}\right|_{k_{0}}$ until $t=2T$. This sequence repeats, returning the system to its original state at every $2T$, characterized by a subharmonic response at $\pm k_{0}$ with a frequency half that of the driving frequency~\cite{inttc:chandra:2024}. If observations were made at multiples of $T$, the Hamiltonian would seem stationary as the system cyclically forms and disintegrates a Cooper pair at $\pm k_{0}$. This behavior defines a DTC phase in which the temporal $Z_2$ symmetry is spontaneously broken.

To ensure that the phase remains rigid and impervious to microscopic changes within the system parameters, it is imperative that equations~\eqref{eq:tc:criteria} consistently produce a solution within the Hamiltonian parameter space $\left|H_{1}\right|_{k}$ for any arbitrary selection of drive parameters $g_{0}, \omega$. Consequently, this phase will rapidly deteriorate if $\left|H_{1}\right|_{k}$ is characterized by the absence of additional parameters, given that the two equations will only produce a singular solution of $\omega$ for any value of $g_{0}$, leading to a rapid degradation of the subharmonic if $\omega$ slightly deviates from this value. However, the inclusion of one or more additional parameters will ensure the existence of infinite solutions for any selection of $g_{0}$, thus merely causing $k_{0}$ to change minimally should $\omega$ be adjusted.  The transverse field Ising model (TFIM) in a spin chain of length $N$ with nearest-neighbor Heisenberg exchange interactions showcases a single-parameter scenario. The Hamiltonian
\begin{equation}
H =  2\sum^{N-1}_{i=0} \left( g_{0} \hat{S}^x_{i}+ \hat{S}^z_i \hat{S}^z_{i+1}\right),
\end{equation}
is described by vector operators $\bm{\hat{S}}_i$, which capture the quantum mechanics of spin-$1/2$ particles or qubits, modeling systems like quantum simulators~\cite{georgescu:2014,you:2005} with Josephson junctions~\cite{kockum:2019} and gas microscopes with trapped ultra-cold ions~\cite{blatt:2012,ultracold:bissbort:2013,ott:2016} to mesoscopic ferromagnets~\cite{nguyen:2017, ostman:2018} and quantum cellular automata~\cite{cowburn:2000, imre:2006}. The transverse field $g_{0}$ breaks the spatial $Z_2$ symmetry of the exchange interaction $\hat{S}^z_i\hat{S}^z_{i+1}$ and is expressed relative to the exchange energy, scaled to unity. Using a Jordan-Wigner transformation~\cite{jordan:1928, lieb:schultz:mattis:1961, pfeuty:1970, tsvelik:2003,derzkho:2008, ising:mbeng:2024}, this Hamiltonian transforms to $\left|H_{1}\right|_{k}$ with $b_{k} = \cos{k}, \Delta_{k} = \sin{k}$ and $g_{0}$ as the sole parameter. Subharmonic patterns in $\pm k_{0}$ quickly degrade, but this issue is addressed in~\cite{inttc:chandra:2024} by extending to higher dimensions where $\left|H_{1}\right|_{k}$ emerges from a hexagonal spin network (also known as the Kitaev model~\cite{pachos:2012,hermanns:2018,takagi:2019}) with $b_{\bm{k}} = \cos{k_x} + \cos{k_y}, \Delta_{\bm{k}} = \sin{k_x} + \sin{k_y}$, employing the additional component of the now vectorized momentum $\bm{k} = k_x\hat{x} + k_y\hat{y}$ as a new parameter. We now explore the alternative, which involves the inclusion of more exotic one-dimensional spin-spin interactions, to introduce the required additional parameters without expanding dimensionally. For example, the Hamiltonian
\begin{equation}
    H =  2\sum^{N-1}_{i=0} \left( g_{0} \hat{S}^x_{i}+ \hat{S}^z_i \hat{S}^z_{i+1} + \lambda  \hat{S}^z_{i-1} \hat{S}^x_i \hat{S}^z_{i+1} \right)
\label{eq:hamiltonian:longrange}
\end{equation}
models the additional roles of longer-range interactions. The first term is the transverse field, the second is the nearest-neighbor Heisenberg exchange, and the third is a 3-body interaction with strength $\lambda$ that can be used to model longer-range interactions that arise in quantum wires~\cite{3body:niu:2012,sarkar:2018}.
The Jordan-Wigner transformation yields $\left|H_{1}\right|_{k}$ with $b_{k} = \cos{k} + \lambda\cos{2k}, \Delta_{k} = \sin{k} + \lambda\sin{2k}$.
The Hamiltonian $\left|H_{1}\right|_{k}$ now contains two parameters, $g_{0}$ and $\lambda$, which can be adjusted independently. The Bogolon energies
$E_{k}(g_{0}, \lambda) = \sqrt{\left[g_{0}-b_{k}(\lambda)\right]^2 + \Delta^2_{k}(\lambda)}$ for the pair $k, -k$ are the eigenvalues, as can be inferred by comparing them with Eq.~\eqref{eq:BogHamiltonian}. Detailed profiles of the diagram of the equilibrium phase are discussed in~\cite{sarkar:2018, kumar:2021}. In the driven case, the additional parameter $\lambda$ allows the introduction of new solutions to equations~\eqref{eq:tc:criteria} for any given $g_{0}, \omega$, ensuring that the time crystal phase remains rigid.

The dynamics induced by the driving protocol that alternates between $\left|H_{1}\right|_{k}$ and $\left|H_{2}\right|_{k}$ can now be studied to obtain the time crystal phase in a 2-parameter space of $g_{0}, \lambda$, whose phase profile we will explore in Sect.~\ref{sec:rigidity}.
Quantum Floquet theory provides a powerful framework for analyzing this dynamics due to the presence of the time-periodic drive with period $T$ and frequency $\omega$ (see~\cite{ising:mbeng:2024, floquet:rahav:2003} and the references therein). If the quantum system is closed, then the central object is the propagator for multiple periods $U(nT) = \mathcal{T}\exp\left(-{i}\int_{0}^{nT}dt'\;H(t') \right)$, where $\mathcal{T}$ denotes time ordering, and $H(t')$ is the full time-dependent Hamiltonian. As a consequence of Floquet's Theorem, $U(nT)$ can be decomposed as $U(T) = e^{-iK(nT)} \left(e^{-iH_F T}\right)^n$, where $K(t)$ is the micromotion operator~\cite{bukov:2015} with period $T$ satisfying $K(t+T) = K(t)$ and can be set to vanish at $t=nT$ without loss of generality. In addition, $H_F$ is the time-independent Floquet Hamiltonian (also called the effective Hamiltonian). This decomposition allows the stroboscopic dynamics at integer multiples of the driving period to be governed by the simple exponential $U(nT) = \left(e^{-iH_F T}\right)^n$, while the micromotion operator $K(t)$ captures the intra-period oscillations. The Floquet Hamiltonian eigenvalues $\Omega_\alpha$ (quasienergies) are defined modulo $\omega$, and the corresponding Floquet states $\ket{\Psi_\alpha(t)} = e^{-i\Omega_\alpha t}\ket{\Phi_\alpha(t)}$ evolve as Bloch waves in time, where $\ket{\Phi_\alpha(t)}$ are the $T-$periodic Floquet modes.

Floquet's Theorem can be applied to each distinct local $k, -k$ sector within this integrable system. Here, the time evolution is determined by the sector Hamiltonian $H_{k}(t)$, as specified in Eq.~\eqref{eq:quadratic}. The respective alternating Hamiltonians $\left\vert H_{1,2} \right\vert_{k}$ in Eqs.~\eqref{eq:H12k} can be associated with sector Hamiltonians $H^{(1)}_{k}(g_{0},\lambda)$ and $H^{(2)}_{k}(g_{1})$ through the Nambu spinor representation.
\begin{align}
    H^{(1)}_{k}(g_{0},\lambda) &= E_{k}(g_{0}, \lambda)\;\bm{n}_{k}(g_{0}, \lambda)\cdot\bm{\tau} \nonumber\\
    H^{(2)}_{k}(g_{1})&=g_{1}\tau_3.
\label{eq:localhamilts}
\end{align}
Here, $\bm{n}_{k} (g_{0}, \lambda)$ is a unit vector on the surface of a Bloch sphere, given by the equation
\begin{align}
    \bm{n}_{k} (g_{0}, \lambda) &\equiv n_{1k}(g_{0}, \lambda)\hat{x} + n_{3k}(g_{0}, \lambda)\hat{z}\nonumber\\
        &=\frac{\Delta_{k}(\lambda) \hat{x} + \left[g_{0}-b_{k}(\lambda)\right]\hat{z}}{E_{k}}.
\end{align}
Applying Floquet's theorem yields the propagator in each sector at times $t=nT$ to be
\begin{equation}
    U_{k}(nT) \equiv \bigg[e^{-iH^{(2)}_{k}(g_{1})T/2}\; e^{-iH^{(1)}_{k}(g_{0},\lambda)T/2}\bigg]^n = \left(e^{-iH^F_{k}T}\right)^n,
\label{eq:localhf}
\end{equation}
where $H^F_{k}$ is the Floquet Hamiltonian corresponding to $H_{k}(t)$. Now, if we denote the eigenvalues by $\pm\theta_{k}/T$, then we can write
\begin{equation}
    H^F_{k} \equiv \frac{1}{T}\;\theta_{k}(g_{0}, \omega, \lambda)\;\bm{h}_{k}(g_{0}, \omega, \lambda)\cdot\bm{\tau},
\label{eq:floq:hamilt:spinor}
\end{equation}
where $\bm{h}_{k} (g_{0}, \lambda, g_{1})$ is the unit vector on the surface of the Bloch sphere that describes the Floquet Hamiltonian in this particular sector. Finally, substituting the RHS of Eq.~\eqref{eq:floq:hamilt:spinor} into the RHS of Eq.~\eqref{eq:localhf} and comparing the traces on both sides after substituting Eqs.~\eqref{eq:localhamilts} into the LHS yields $\cos{\theta_{k}} = \Re{A_{k}}$, where
\begin{multline}
    A_{k}(g_{0}, \omega, \lambda) =e^{-i g_{1} T / 2}\Bigg\{\cos \left[\frac{E_{k}(g_{0}, \lambda)\ T}{2}\right]\\
        -i n_{3 k}(g_{0}, \lambda)\ \sin \left[\frac{E_{k}(g_{0}, \lambda)\ T}{2}\right]\Bigg\}.
\label{eq:Ak}
\end{multline}
From Eqn.~\eqref{eq:Ak}, it can be seen that gaplessness occurs when $\Re{A_{k}}=1$, or 
\begin{multline}
    \cos{\bigg[\frac{g_{1}T}{2}\bigg]}\cos{\bigg[\frac{E_{k_g}(g_{0},\lambda)T}{2}\bigg]}\\
        - \bigg[\frac{g_{0}-b_{k_g}(\lambda)}{E_{k_g}(g_{0},\lambda)}\bigg]\;\sin{\bigg[\frac{g_{1}T}{2}\bigg]}\sin{\bigg[\frac{E_{k_g}(g_{0},\lambda)T}{2}\bigg]} = 1.
\label{eq:gaplessness}
\end{multline}
\begin{figure*}[hbtp]
    \includegraphics[height=0.7\textheight, keepaspectratio]{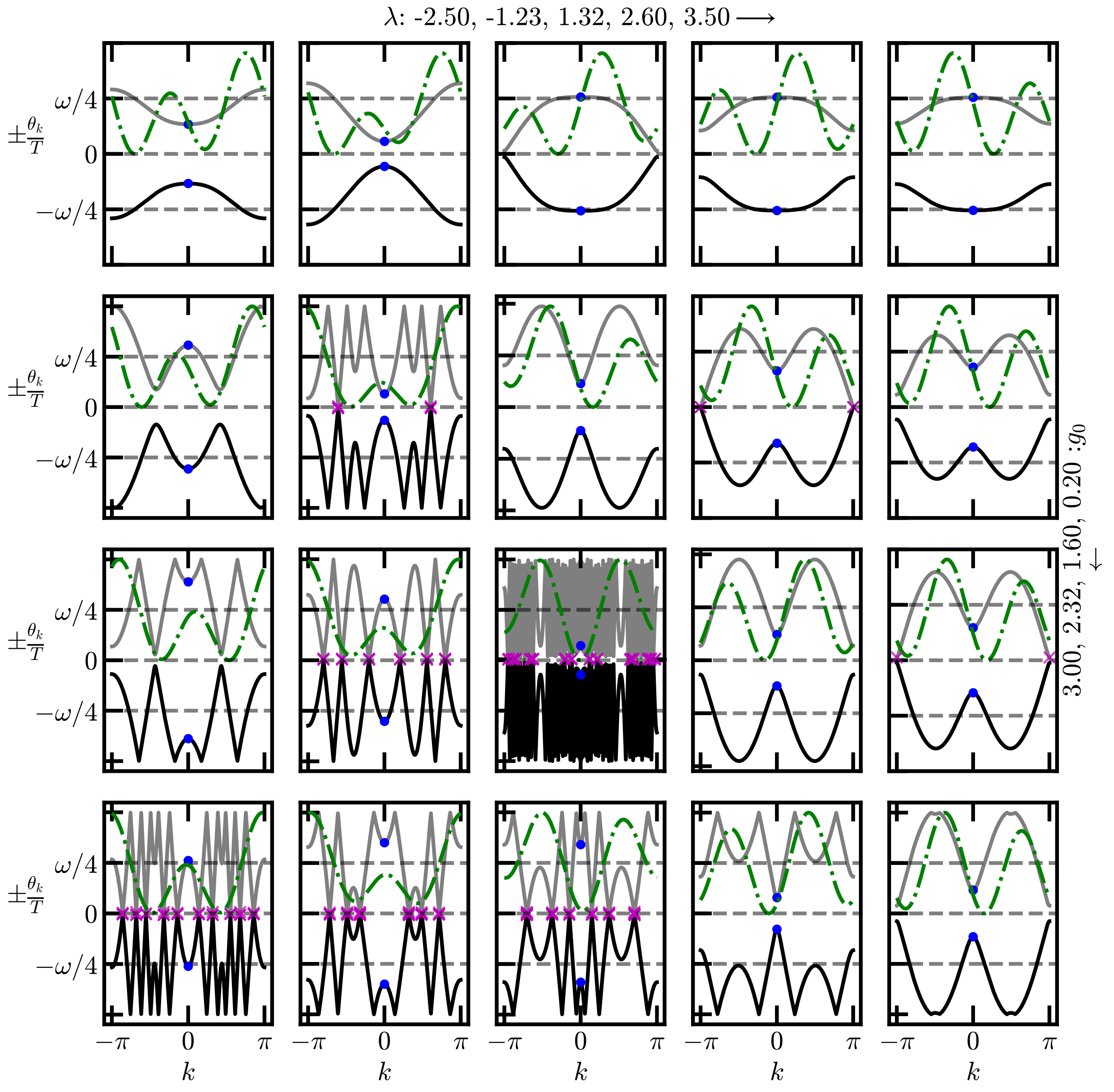}
    \caption{Floquet quasienergies $\pm\theta_{k}$ (in units of time period $T$) derived from Eqn.~\eqref{eq:Ak}. Each row corresponds to a fixed value of $g_{0}$, while different values of $\lambda$ are systematically displayed across columns. The values of $g_{0}$ and $\lambda$ are clearly indicated above the topmost panels and after the rightmost panels, respectively. The values of $\pm \theta_{k}$ at the high-symmetry gapless point (non high-symmetry gapless points) are indicated by blue (magenta) colored dots. Additionally, the parameter $g_{1}=2\omega$, and $\omega$ has been optimized according to the cost function in Eqn.~\eqref{eq:costfunction}. Finally, the cost function itself has been plotted using green dot-dashed lines, with the ordinate values on the right axis (arbitrary units) in each panel. A subharmonic response exists only when a minimum in the cost function touches $0$, and the quasienergy curves intersect $\pm\omega/4$ at the same value of $k$.}
\label{fig:quasienergies}
\end{figure*}
Additionally, the subharmonic mode appears in the quasienergy spectrum at momenta $k_0$ when Eq.~\eqref{eq:tc:criteria} is satisfied, giving $\theta_{k_0}/T=\pm\omega/4$. Finally, we define \emph{equilibrium resonance} by the condition $g_1=2n\omega$ (with integer $n$), Eq.~\eqref{eq:gaplessness} then reduces to $\cos\big[E_{k_g}(g_0,\lambda)T/2\big]=1$, i.e. $E_{k_g}(g_0,\lambda)=2n\omega$, which for $n=0$ coincides with the equilibrium gapless points. 

\section{Rigidity and Phase Diagram of the Time Crystal}
\label{sec:rigidity}
The rigidity of the time crystal phase can be enforced by the additional parameter $\lambda$ in the Hamiltonian $\left|H_{1}\right|_{k}$, which allows the existence of solutions to equations~\eqref{eq:tc:criteria} for any arbitrary choice of $g_{0}, \omega$. When the integrable time-crystal paradigm is implemented in spin-chain systems, the existence of bounds on the right-hand side of these equations implies the presence of regions in parameter space where real solutions cease to exist. Crossing these boundaries induces phase transitions from the DTC to an alternative phase, as has been observed in similar systems.~\cite{inttc:chandra:2024,kumar:2021} This section explores the phase diagram of the time crystal in the two-parameter space of $g_{0}, \lambda$ for certain values of $g_{1}, \omega$. The phase diagram is constructed by analyzing the quasienergy spectrum $\pm\theta_{k}$ derived from Eq.~\eqref{eq:Ak}, as well as the DTC conditions in Eqs.~\eqref{eq:tc:criteria} and identifying regions where stable subharmonic modes exist.

\subsection*{Analytical Phase Diagram}
We apply the DTC conditions derived in Sect.~\ref{sec:isingchain} to the long-range spin chain parameterised by $(g_0,\lambda)$. The subharmonic (period-doubled) requirement for a momentum sector $k$ is as follows.
\begin{align}
    f(k;g_0,\lambda) &= g_0 - b_k(\lambda) = 0,\nonumber\\
    \omega(k;\lambda) &= 2\Delta_k(\lambda),
\label{eq:spinchain:dtc}
\end{align}
with
\begin{equation}
    b_k(\lambda)=\cos k + \lambda\cos 2k,\qquad
    \Delta_k(\lambda)=\sin k + \lambda\sin 2k.
\end{equation}
For fixed $\lambda$ the first equation admits real solutions only when
\begin{equation}
    g_0\in B_\lambda\equiv\big[\min_k b_k(\lambda),\,\max_k b_k(\lambda)\big],
    \label{eq:blambda}
\end{equation}
so that the admissible region in the $(g_0,\lambda)$–plane is the filled set swept out by $b_k(\lambda)$.

To obtain $B_\lambda$ explicitly, it is convenient to set $x=\cos k$, so
\begin{equation}
    b(x;\lambda) = -2\lambda x^2 + x + \lambda,\qquad x\in[-1,1].
\end{equation}
The critical points satisfy $b'(x)=0$, giving $x^\ast=(4\lambda)^{-1}$, which lies in $[-1,1]$ only for $|\lambda|\ge 1/4$. Evaluating $b$ at the boundary and at $x^\ast$ yields
\begin{equation*}
    b(\pm1)=\pm1-\lambda,\qquad b(x^\ast)=\lambda+\frac{1}{8\lambda}\quad(\abs{\lambda}\ge 1/4).
\end{equation*}
Hence,
\begin{itemize}
    \item for $|\lambda|<1/4$ there is no interior extremum and
    \[
        B_\lambda=[-1-\lambda,\;1-\lambda],
    \]
    \item for $|\lambda|\ge 1/4$ the interval endpoints are the extrema among $\{b(-1),\,b(1),\,b(x^\ast)\}$.
\end{itemize}

The Floquet quasienergy gaplessness condition (Eq.~\eqref{eq:gaplessness}) can be analyzed in parallel. At the high-symmetry point $k=0$ (where $\Delta_{0}=0$) Eq.~\eqref{eq:gaplessness} reduces to a family of planes in the parameter space,
\begin{equation}
    g_0 = \lambda - g_1 \pm 1 - 2n\omega,\qquad n\in\mathbb{Z},
\label{eq:gapless:hs}
\end{equation}
whereas non-‐high‑symmetry solutions (non-trivial roots of $\Delta_k=0$) produce the complementary set
\begin{equation}
    g_0 = -\lambda - g_1 \pm 1 + 2n\omega.
\label{eq:gapless:nohs}
\end{equation}
Under the equilibrium–resonance condition $g_1 = 2m\omega$, the Floquet gapless points collapse onto the equilibrium gapless manifold specified by $E_k(g_0,\lambda) = 2n\omega$ and consequently closely track the boundaries of $B_\lambda$. Deviations occur solely due to the $\mathcal{O}(1/\lambda)$ contribution in $b(x^\ast)$ for small $|\lambda|$, such that the boundary of the $B_\lambda$ manifold asymptotically converges to the equilibrium gapless manifold.

Taken together, these analytic results provide a compact criterion for the existence of subharmonic solutions: a given $(g_0,\lambda)$ admits a DTC sector whenever $g_0\in B_\lambda$ and the corresponding Floquet gap is open at the momentum solving Eq.~\eqref{eq:spinchain:dtc}. In contrast, crossing the gapless planes in Eqs.~\eqref{eq:gapless:hs}–\eqref{eq:gapless:nohs} signals the closing of the protecting quasienergy gap and marks the approximate boundary between the DTC and Floquet-paramagnetic regimes.

We now examine the structure of the time-crystal phase in greater detail. In Ref.~\cite{inttc:chandra:2024}, it was shown that the line $\lambda=0$ cannot host a stable time crystal: with only a single tunable parameter in $\left|H_1\right|_k$, there exists at most one drive frequency $\omega$ satisfying the DTC conditions for a given $g_0$, rendering the subharmonic response highly sensitive to small detunings of $\omega$. Even for $\lambda\neq 0$, some residual sensitivity is expected. One practical strategy to mitigate this is to employ a hybrid quantum--classical variational loop in which a classical optimizer measures the relevant observables, evaluates a cost function, and updates the control parameters $(k,\omega)$ accordingly. Such closed-loop control can stabilize parameter regions that would otherwise be fragile under fixed drives~\cite{cerezo:2021}.

Specifically, we optimize the pair $(k,\omega)$ by minimizing the cost function
\begin{equation}
    F(k,\omega) = \big[g_{0}-b_{k}(\lambda)\big]^2 + \bigg[\frac{\omega}{2}-\Delta_{k}(\lambda)\bigg]^2.
\label{eq:costfunction}
\end{equation}
Minimization over $\omega$ is immediate and yields $\omega = 2\Delta_k(\lambda)$, reducing the problem to a single-variable objective,
\begin{equation*}
  F(k) = \big[g_0 - b_k(\lambda)\big]^2.
\end{equation*}
Any root of $f(k;g_0,\lambda)=0$ is a global minimizer. Generically, this equation admits two solutions $k=\pm k_0$; when $|\lambda|\ge 1/4$, up to four solutions may exist. The map $(g_0,\lambda)\mapsto k$ is therefore multivalued, and a local optimizer initialized stochastically will converge to the minimum within its basin of attraction. Small perturbations in $(g_0,\lambda)$ or in the initial guess can cause the optimizer to jump between equivalent minima, producing apparent discontinuities in the optimized $k$. This effect is most pronounced near $g_0 = -\lambda$, where $k=\pm\pi/2$ are exact solutions that produce $\omega = \pm 1$ at the same cost. Below this line, the solver alternates between these degenerate branches. Sufficiently far above, minima near $k\approx 0$ or $\pi$ dominate, and a single branch yields a well-defined DTC subharmonic whenever the cost vanishes.

\begin{figure}[t!]
    \centering
    \includegraphics[width=\linewidth, keepaspectratio]{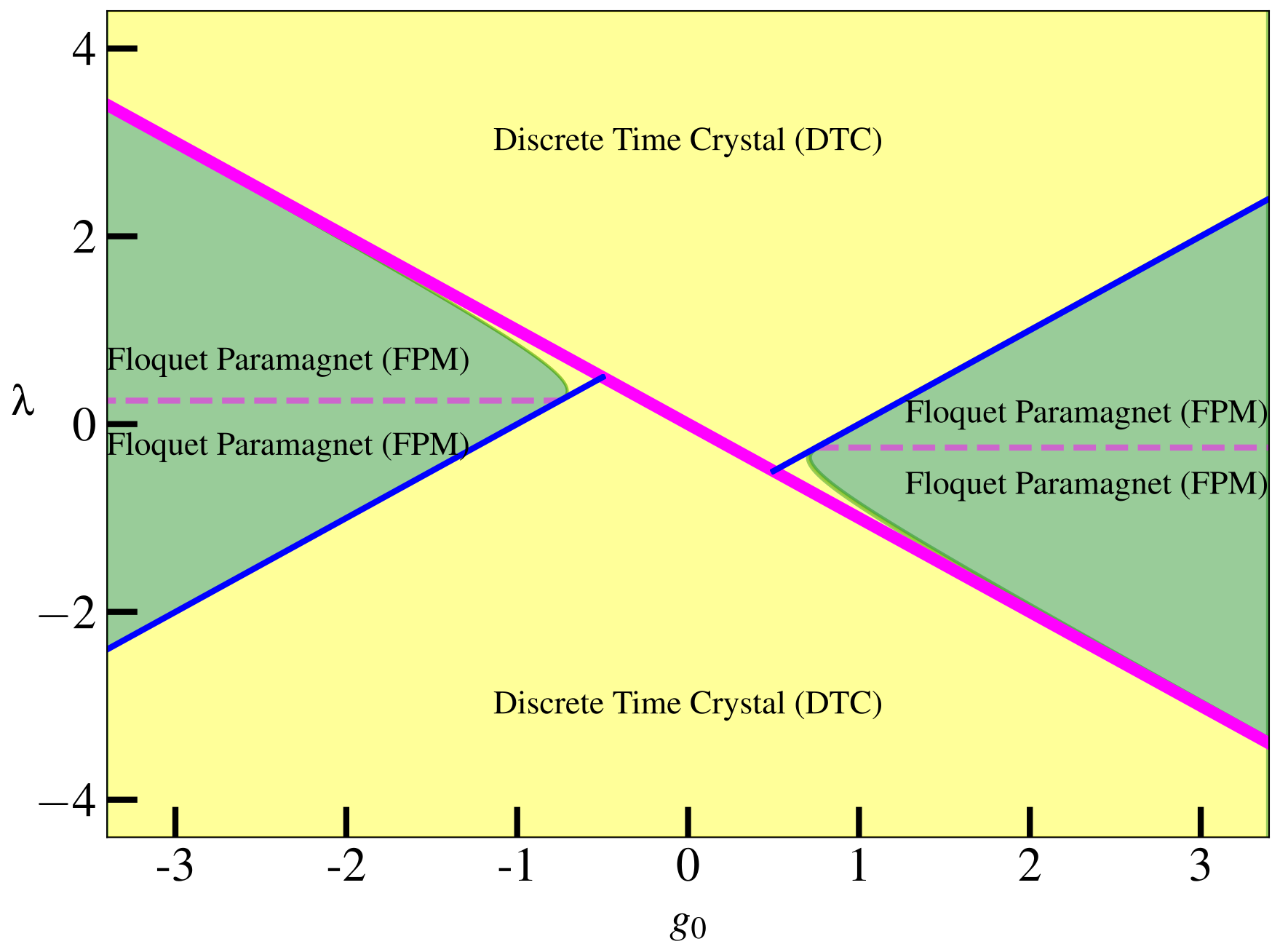}
    \caption{Phase Diagram of the driven chain. The filled region $B_\lambda$ (shaded yellow/light yellow and labeled "Discrete Time-Crystal (DTC)") in the $(g_0,\lambda)$ plane marks parameters that admit subharmonic solutions. Regions outside this area (shaded green and labeled "Floquet paramagnet (FPM)") do not support subharmonic roots. Solid lines show the analytically determined Floquet gapless points [Eqs.~\eqref{eq:gapless:hs} and~\eqref{eq:gapless:nohs}] for $g_1=2\omega$, with blue (magenta) lines indicating gaplessness at HS (non-HS) momenta. These gapless curves closely track the boundaries of $B_\lambda$ and thus serve as approximate phase boundaries between the DTC and FPM regimes. Finally, the horizontal dashed lines at $\abs{\lambda}=1/4$ divide the two regions of $B_\lambda$ with one or no interior extrema.}
\label{fig:phase:space}
\end{figure}

Figure~\ref{fig:phase:space} displays the filled set $B_\lambda$ in the $(g_0,\lambda)$ plane together with the Floquet gapless points for $g_1=2\omega$. The shaded regions delineate parameters that support subharmonic solutions (DTC) and those that do not (FPM), providing a transparent phase diagram of the driven system. The nonequilibrium phase structure can also be read directly from Fig.~\ref{fig:quasienergies}, which shows the Floquet quasienergy spectra $\pm\theta_k$ for representative pairs $(g_0,\lambda)$ alongside the corresponding cost-function profiles. Each row corresponds to a fixed $g_0$, with $\lambda$ varying between columns. The dashed horizontal lines at $\theta_k/T=0,\pm\omega/4$ serve as reference scales. Blue dots and magenta crosses mark gapless points at high-symmetry (HS) and non-HS momenta, respectively; green dot-dashed lines show the cost function (right axis). In the FPM phase, the cost function does not vanish for any $k\neq 0$; if a zero occurs, it does so at $k=0$ where $\omega=0$, which does not produce no subharmonic. In the DTC regime, on the contrary, the cost function attains a zero minimum at some $k_0\ne 0$, and the quasienergy curves cross $\pm\omega/4$ at the same $k_0$, confirming the presence of a subharmonic mode.

The role of the Floquet gap as a diagnostic of phase transitions is equally apparent in Fig.~\ref{fig:quasienergies}: subharmonic solutions emerge as soon as the Floquet gapless points are crossed and the protecting gaps open. The gapless points therefore partition the $(g_0,\lambda)$ plane into regions that support or do not support stable subharmonic solutions. The admissible region closely coincides with $B_\lambda$, while the complementary region exhibits paramagnetic behavior. Consequently, the analytically obtained gapless points serve as approximate phase boundaries separating the DTC and FPM regimes.
\begin{figure*}[t!]
    \centering
    \includegraphics[width=0.75 \textheight, keepaspectratio]{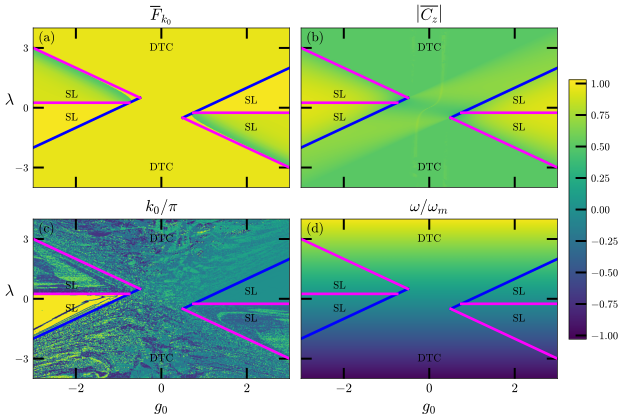}
    \caption{Density plot of the long-time average (strobed at even multiples of $T$)  of the fidelity $\overline{F}_{k_{0}}$ at the optimal momentum $k_{0}$ (panel a), and the correlations $\overline{C_z}$ (panel b), strobed at integer multiples of $T$. In panel b, the FBZ was discretized into $N=1000$ equally spaced points between $-\pi$ and $\pi$, and $k_{0}$ explicitly included in the chosen sum. The time average is performed over $10^4$ sets of $2T$-intervals, where $T=2\pi/\omega$. The parameter space is chosen to be $g_{0},\lambda$. For each point, the value of $\omega$ is optimized using the trust-region method to minimize the cost function in Eqn.~\eqref{eq:costfunction}. The trust region predictions for $k_0$ ($\omega$) are plotted in panels c (d). The parameter $g_{1}$ is set to $2\omega$, the equilibrium resonance condition described in the text. The solid lines indicate the gapless points in the Floquet quasienergy spectrum (Eqs.~\eqref{eq:gapless:hs} and~\eqref{eq:gapless:nohs}), where the lines that correspond to gaplessness in the HS (non-HS) points are colored blue (magenta). The gapless points divide the parameter space into DTC and FPM phases, as labeled in the figure.}
\label{fig:fidelity_long_time_avg}
\end{figure*}
\subsection*{Numerical Phase Diagram}
We complement the analytical phase diagram with a numerical survey of the parameter plane $(g_0,\lambda)$. This approach simultaneously provides an independent validation of the analytical predictions and offers a systematic means to identify additional physics that may not be captured by purely analytical considerations. Furthermore, numerical evaluations of the matrix elements of suitably chosen observables will elucidate the distinctions between the two phases and will characterize the emergence of off-diagonal long-range order in the DTC phase.

At each point on the $(g_0,\lambda)$ grid, we numerically optimize subharmonic conditions to identify candidate pairs $(k_0,\omega)$ and evaluate stroboscopic diagnostics such as long-term average fidelity and two-point correlations. These calculations, implemented with robust minimization and spectral analysis routines, yield a comprehensive Floquet phase map that can be compared directly with the analytically obtained gapless points.

For every pair $(g_{0},\lambda)$, we look for $(k_{0},\omega)$ that minimizes the cost function in Eqn.~\eqref{eq:costfunction}. Minimization is performed using a trust-region method from the SciPy library~\cite{scipy:nmeth:2022}, with numerical derivatives supplied by numdifftools~\cite{numdifftools}. An exact solution of Eqs.~\eqref{eq:spinchain:dtc} corresponds to a zero minimum of the cost, in which case the sector fidelity
\begin{equation}
    F_{k_{0}}(2nT)\equiv\big|\langle\psi_{k_{0}}(0)\mid\psi_{k_{0}}(2nT)\rangle\big|^2
\end{equation}
reaches unity. When no exact root exists, the minimizer returns a nonzero value and the stroboscopic fidelity at $2nT$ falls below unity, signaling melting of the time-crystalline order. Consequently, the long-time average $\overline{F}_{k_{0}}(2nT)$ (arithmetic mean over many even periods) is used as a primary diagnostic for the DTC phase; equivalently, $1-\overline{F}_{k_{0}}(2nT)$ can be interpreted as the defect density in the $k_0$ sector~\cite{roy:2013}. All fidelity and dynamical observables reported here are computed using QuTiP~\cite{qutip5}.

Figure~\ref{fig:fidelity_long_time_avg} (left panel) shows density plots of $\overline{F}_{k_{0}}(2nT)$, where $\omega$ is chosen at each point to minimize the cost function using the trust-region method; the overline denotes an average over long $n$. In addition, the parameter $g_1$ has been set to $2\omega$, the choice of equilibrium-resonance. 
By default, we choose to initialize the system in the fully polarized state $\ket{\psi(0)} = \bigotimes_i \ket{\downarrow}_i$, which is a ground state of the Hamiltonian when $g_{0}$ is large and negative. Thus, this figure shows the Floquet phase diagram in the $(g_0,\lambda)$ plane.  The time averages are taken over two-period intervals $10^4$ ($T=2\pi/\omega$). The regions with $\overline{F}_{k_{0}}\approx 1$ correspond to a time crystal (DTC): the system supports persistent stroboscopic subharmonic oscillations pinned at optimal momentum $k_0$. 
The lower panels of Fig.~\ref{fig:fidelity_long_time_avg} display the predictions of the trust-region algorithm for optimal $\omega$ and $k_0$ in the parameter space. These plots align closely with the analytical predictions in Sect.~\ref{sec:rigidity}, confirming the validity of the optimization approach. The time crystal phase (high $\overline{F}_{k_{0}}$) appears in the same regions where analytical DTC conditions are satisfied, while the FPM regime (variable $\overline{F}_{k_{0}}$) appears where no exact roots exist, except at the high-symmetry point $k_0=0$, where $\omega$ is trivially zero (see the bottom left panel of Fig.~\ref{fig:fidelity_long_time_avg}), yielding spurious unit fidelity despite being in the FPM phase. The solid lines overlaying the plots are the analytically computed Floquet gapless points: blue lines indicate gaplessness at high-symmetry (HS) momenta and magenta lines indicate gaplessness at non-HS momenta. These gapless lines track the actual phase boundaries, with the $g_0=-\lambda$ line converging to the phase boundary (given by the edges of the manifold $B_\lambda$ in Eq.~\ref{eq:blambda}) in the asymptotic limit of large $\lambda$. Thus, when a quasienergy gap is open, a subharmonic can be pinned, and the DTC is stable; when that gap closes, the subharmonic can disappear, and the system falls into the FPM phase. In short, the existence and robustness of the time-crystalline phase are witnessed directly by where and how the Floquet quasienergy gaps open in momentum space. In addition, the lower panels of Fig.~\ref{fig:fidelity_long_time_avg} show that the optimized $\omega$  varies smoothly throughout the phase diagram. This shows that tiny deviations of the actual drive frequency from its optimal value lead only to tiny changes in subharmonic momentum $k_0$. In the continuum limit ($N\to\infty$), this merely causes a slight displacement of the subharmonic response within the FBZ, rather than its complete destruction. A detector capable of resolving momentum-space dynamics would consistently observe a subharmonic at or close to $k_0$, even when the drive frequency has small inaccuracies. Consequently, the smooth dependence signifies a Discrete Time Crystal that is robust against small perturbations in the thermodynamic (continuum) limit. 
\begin{figure*}[ht!]
    \centering
    \includegraphics[width=\linewidth]{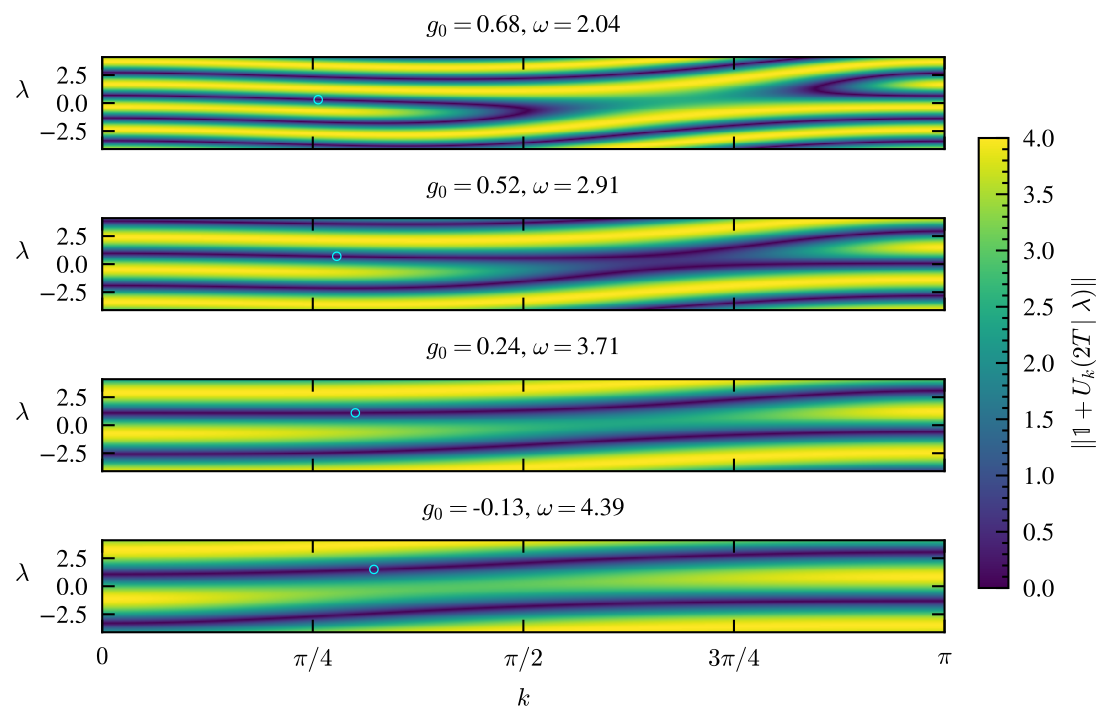}
    \caption{Plots of $\norm{\mathds{1}+ U_k(2T)}$, the trace of the unit-displaced propagator, in $\lambda, k$ space, computed from QuTiP simulations. The fixed values of $g_0, \omega$ (indicated in the title of each panel) are chosen from a small cluster of values, and the parameter $g_1=2\omega$ is set to the equilibrium resonance condition. The driving frequency $\omega$ is not optimized. Nonetheless, the theoretically optimum $k_0, \lambda$ point (where the subharmonic criteria in Eq.~\eqref{eq:tc:criteria} are exactly met) is circled in cyan in each panel.}
    \label{fig:op:norm}
\end{figure*}
The argument for rigidity is further supported by examining the full momentum-space dynamics at a fixed, non-optimized driving frequency, while varying other system parameters. Figure~\ref{fig:op:norm} shows density plots of the trace-norm of the unit-displaced propagator, $\norm{\mathds{1}+ U_k(2T)}$ (with $U_k(2T)$ obtained from Eq.~\eqref{eq:localhf}), for fixed $g_0,\omega$ in the $k,\lambda$ plane. This norm can vanish only at $k=k_0$, where $U_{k_0}(2T)=-\mathds{1}$, indicating the emergence of a subharmonic at that momentum. Using the operator norm is advantageous because the result is entirely independent of initial conditions. The parameter choices correspond to a generic set of nearby points in the $(g_0,\omega)$ space, representing small variations. The theoretical $k_0,\lambda$ location where the subharmonic criteria in Eqs.~\eqref{eq:tc:criteria} are satisfied is marked. These plots clearly show that small parameter changes merely shift the subharmonic feature slightly to neighboring momenta within the FBZ. This behavior is consistent with the smooth dependence of $k_0$ on the parameters observed in the lower panels of Fig.~\ref{fig:fidelity_long_time_avg}, and further confirms the robustness of the time-crystalline phase against small perturbations in the drive frequency and other system parameters.

More physically measurable distinctions between these regions can be drawn by examining the full many-body temporal correlations. In the Heisenberg picture, these are given by
\begin{align}
    C_z(nT) &= \sum_i \expval{\sigma^z_i(0)\; \sigma^z_i(nT)} \nonumber \\
        &= \sum_i \expval{\sigma^z_i e^{iH_F\;nT} \sigma^z_ie^{-iH_F\;nT}},
\end{align}
where $H_F$ is the full many-body Floquet Hamiltonian. The operator $\sigma^z_i$ flips the spin at the site $i$, creating a pair of Bogolons with momenta $\pm k$ in each sector. The time evolution under $H_F$ then causes these Bogolons to evolve independently, as the Hamiltonian is integrable. The correlation function $C_z(n)$ can thus be expressed as a sum of the contributions from each momentum sector:
\begin{multline}
    C_z(nT)  = \frac{1}{N}\sum_{k} \bra{\psi_{k}(0)}
            {\exp{iH^F_{k}\cdot nT}\;\tau_3 \\
            \exp{-iH^F_{k}\cdot nT}}\ket{\psi_{k}(0)}
\label{eq:correlations:kspace}
\end{multline}
where $H^F_{k}$ is the Floquet Hamiltonian in the $k,-k$ sector shown in Eqn.~\eqref{eq:floq:hamilt:spinor}, and $\ket{\psi_{k}(0)} = \ket{0}$ is the vacuum state in that sector.

The long-time average of $\overline{C_z}$, is plotted in the right panel of Fig.~\ref{fig:fidelity_long_time_avg} and shows a phase manifold that is similar to that revealed by the fidelity analysis. The DTC region has a value of $\overline{C_z}\approx 0.5$, while the FTC region shows a slight increase in $\overline{C_z}$. In the FPM regions, the correlations show a marked increase from nearly half at the phase boundaries to nearly unity in regions with sufficiently large $\abs{\lambda}$.  
\subsection*{Off-Diagonal Long-Range Order}
The nomenclature ``Floquet Paramagnet'' (FPM) and ``Discrete Time Crystal'' (DTC) is further justified by looking at the scaling behavior of the \emph{temporal} off-diagonal long-range order (ODLRO) encoded in the antidiagonal elements of the temporal covariance matrix.  Formally, we define the $N\times N$ temporal covariance matrix
\begin{align}
    \rho_{nm} &= \expval{\tau_3(k_0, n)\,\tau_3(k_0, m)}{\psi_{k_0}(0)},\label{eq:temporal_cov}\\
    \tau_3(k_0, n) &\equiv U^\dagger_{k_0}(2nT)\,\tau_3\,U^{\;}_{k_0}(2nT),\nonumber
\end{align}
where $U_{k_0}(T)$ is the Floquet propagator on the optimized
$(k_0,\omega)$ given by eq.~\eqref{eq:localhf}, and the stroboscopic time index $n$ runs from $N/2$ to $N$.  The anti-diagonal entries $\rho_{n,N-n}$ probe the maximal temporal
separation between two spin-flip events. This quantity is an exact temporal analog of the off-diagonal spatial correlator $\expval{ a^\dagger_n  a^\dagger_m a_n a_m}$ at
large spatial separation $|n-m|$ whose nonvanishing thermodynamic signals spatial ODLRO in superfluids and magnetic condensates~\cite{yang:1962,penrose:onsager:1956}.
By averaging over these maximally separated pairs,
\begin{equation}
    \overline{\abs{\rho_{n,N-n}}}
    \;=\;
    \frac{1}{N}\sum_{n=N/2}^{N}
    \bigl|\rho_{n,N-n}\bigr|,
\label{eq:odlro_avg}
\end{equation}
we construct a scalar diagnostic that is of the order of unity whenever
long-range temporal order persists, and decays to zero in its absence.

\begin{figure}
    \centering
    \includegraphics[width=\linewidth]{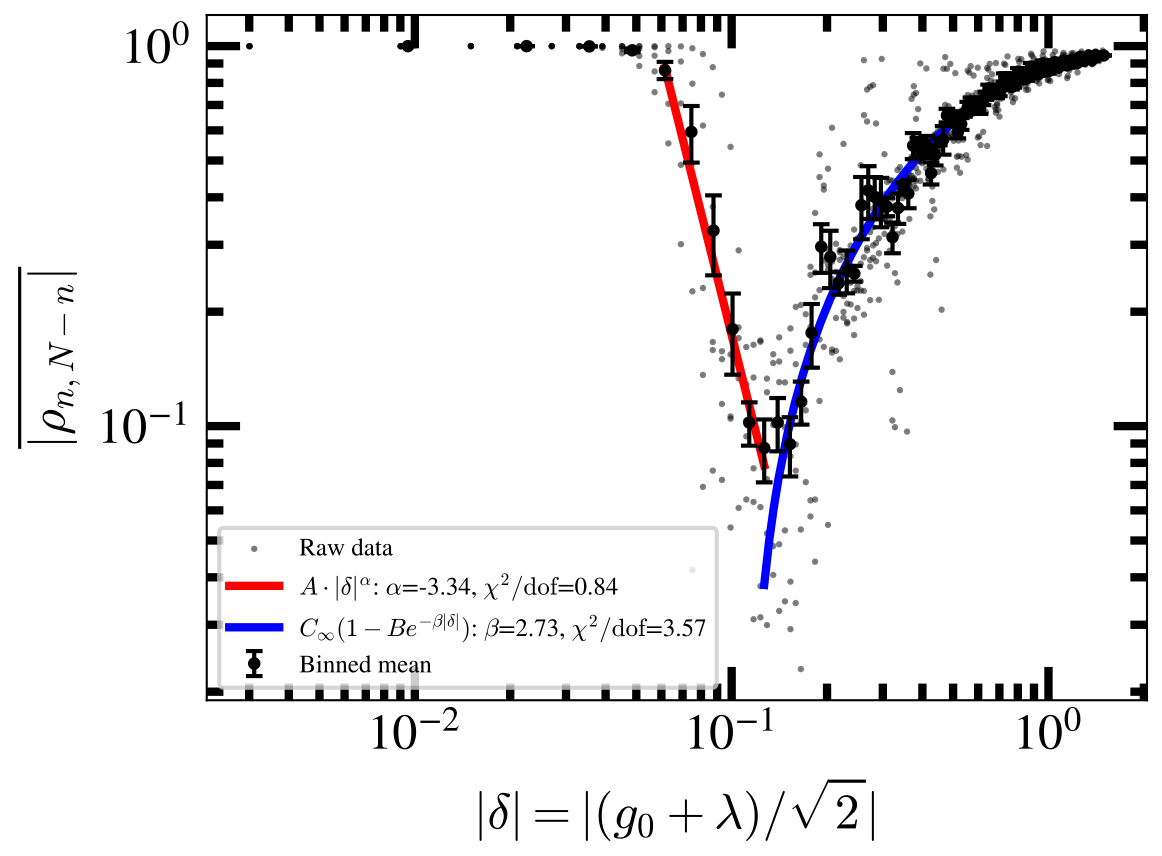}
    \caption{Log-log plots of temporal correlations averaged over the anti-diagonal elements of the temporal covariance matrix $\rho_{nm}$, defined in Eq.~\eqref{eq:temporal_cov}. The panel plots $\overline{\abs{\rho_{n, N-n}}}$ defined in Eq.~\eqref{eq:odlro_avg} as a function of $\abs{\delta}$, where $\delta = (g_0 + \lambda)/\sqrt{2}$ is the Euclidean distance from the line $\lambda = -g_0$. The raw data (small gray dots) is taken from an unordered set in the $g_0,\lambda$ plane; only points with $\delta < 0$ are retained. For each such point, the temporal correlation series is obtained via QuTiP simulations with time steps of $2T$, truncated at $t = NT$ with $N=100$. For each value of $\delta$, all corresponding values of $\overline{\rho}$ from the relevant $g_0,\lambda$ pairs are collected into histograms with $115$ bins; the mean of each histogram is shown as a large black circle, and the error bars represent the standard deviation. In the Floquet Paramagnetic (FPM) regime, these binned means are fitted to the following scaling hypothesis: power-law collapse near criticality (red) and exponential growth in deep-FPM (blue). The fitted parameters and corresponding $\chi^2$ values from the hypothesis tests are given in the legend.}
    \label{fig:odlro}
\end{figure}

Figure~\ref{fig:odlro} plots the diagnostic in Eq.~\eqref{eq:odlro_avg} as a function of $|\delta| = |g_0+\lambda|/\sqrt{2}$, the Euclidean distance in the parameter space from the asymptotic critical line $\lambda = -g_0$, for points in the region $\delta < 0$. Most of this region consists of the FPM phase, although it also includes a small region of the DTC phase due to overlap with the exact $B_\lambda -$manifold described in the introduction to this section (also see Fig.~\ref{fig:phase:space}).

The scaling laws revealed by this analysis furnish a precise temporal
counterpart to the well-known ferromagnet-to-paramagnet transition in the
quantum Ising chain. There, the spatial correlations $\langle\sigma^z_i
\sigma^z_j\rangle$ decay algebraically $\sim|i-j|^{-\eta}$ at
criticality, exponentially $\sim e^{-|i-j|/\xi}$ in the paramagnetic
phase, and saturate to $m_s^2\neq 0$ in the ferromagnetic phase, with
$m_s$ the spontaneous magnetization signaling spatial ODLRO~\cite{pfeuty:1970, lajzerowicz:pfeuty:1975}.  An entirely
parallel structure emerges here in the temporal domain.
\begin{enumerate}[label=(\roman*)]
    \item \textbf{DTC phase}: $\overline{|\rho_{n,N-n}|} = \mathcal{O}(1)$,
        signaling \emph{temporal ODLRO}.  The system remembers its initial
        spin polarization across arbitrarily many drive periods, the
        precise temporal analog of long-range ferromagnetic order.
    \item \textbf{Near the DTC-FPM critical boundary} ($|\delta|\lesssim 1$):
        a power-law collapse on the FPM side
        \begin{equation}
            \overline{|\rho_{n,N-n}|} \sim |\delta|^{\alpha},
        \label{eq:odlro:powerlaw}
        \end{equation}
        with good goodness-of-fit ($\alpha \approx -3.34, \chi^2/\mathrm{dof}\approx 0.84$), mirrors the algebraic decay of
        correlations at a quantum critical point with anomalous dimension
        $\eta$.  The divergence as $\delta\to 0^-$ reflects the
        diverging temporal correlation ``length'' (measured in units of
        the drive period) as the system approaches the phase boundary.
    \item \textbf{Deep FPM} ($|\delta|$ large): exponential saturation to
        a value strictly below unity,
        \begin{equation}
            \overline{|\rho_{n,N-n}|} \sim
            \left[1 - B\,e^{-\beta|\delta|}\right],
        \label{eq:odlro:exp}
        \end{equation}
        with a fitted growth rate $\beta\approx 2.73$, consistent with the exponential suppression of correlations deep in a conventional quantum paramagnetic phase, here manifested in
        the time domain.  The absence of temporal ODLRO is
        the hallmark of a phase that has \emph{not} broken the discrete-time-translation symmetry.
\end{enumerate}

The extracted exponent $\alpha \approx -3.34$ ($\chi^2/\mathrm{dof} \approx 0.84$)  characterizes the critical singularity of the temporal ODLRO at the DTC-FPM  transition; its deviation from standard transverse-field Ising exponents~\cite{sachdev:qpt:2011} suggests that either the composite nature of $\overline{|\rho_{n,N-n}|}$ as an observable 
or the integrability constraints of the Floquet drive modify the effective 
universality class, a question we leave for future investigation.

The crossover between regimes (ii) and (iii) locates the dynamical equivalent of the Ising correlation length $\xi$: within the FPM, this temporal correlation length is finite, just as a paramagnet possesses only short-range spin correlations. In DTC, by contrast, the temporal correlation length diverges (signaled by the saturation of $\overline{|\rho_{n,N-n}|}$ to $\mathcal{O}(1)$), exactly as the correlation length diverges in a ferromagnet.  This triptych of scaling regimes---temporal ODLRO in the DTC, algebraic critical scaling at the boundary, and exponential paramagnetic suppression in the FPM---provides a rigorous temporal-order-parameter justification for both phase names, placing them on the same conceptual foundation as spontaneous symmetry breaking in equilibrium quantum magnets.
\section{Finite-Size Scaling: Methodology and Results}
\label{sec:finitesize}
\subsection{Model and Parameter Selection }
In order to look at finite-size effects, we study the system dynamics numerically for long times at finite sizes. A crucial point to note is the order of limits. In many-body physics, it is well known that the order of limits can significantly affect the results. For example, in the context of Anderson localization in one dimension, if one first takes the limit of the vanishing disorder strength $h\to 0$ and then considers the thermodynamic limit $N\to\infty$, one might incorrectly conclude that all states are delocalized. However, this conclusion is erroneous because the correct order of limits is to take $N\to\infty$ first, followed by $h\to 0$. This ensures that even an infinitesimal amount of disorder can localize states in an infinite system. For a detailed discussion on this matter as it pertains to Many Body Localization, see~\cite{liu:dtc:2023}.

This means that in our scenario, we first select a pair of $g_{0}, \lambda $ and then determine $k_{0}, \omega$. However, this particular $k_{0}$ may not correspond to a reciprocal lattice point except in the limit of infinite size. For a finite system of size $N$, the reciprocal lattice point nearest to $k_{0}$,
denoted $k^R_{0} (N)=2\pi n/N$ (where $n$ varies with $N$), will only approximately optimize the cost function $f$ in Eq.~\eqref{eq:costfunction}. It is expected that, for the $k^R_{0}, -k^R_{0}$ pair, the subharmonic will exhibit distortions with 'beats' (similar to those observed in~\cite{liu:dtc:2023}) at a frequency of $\delta \Omega (N) = k_{0}-k^R_{0} (N)$. This process disrupts the time crystal with a beat period $t_b (N)\sim \left(\delta \Omega (N)\right)^{-1}$. As $N$ becomes very large, $\delta \Omega (N)$ is expected to decrease inversely with $N$, eventually disappearing as $N\to\infty$ for the DTC phase. For the FPM phase, the behavior is expected to be different, with $\delta \Omega (N)$ potentially stabilizing or even increasing with $N$. This would indicate the absence of any robust subharmonic response. Thus, the proposed DTC "resists infinitely at the thermodynamic limit", although the scaling of melting time is linear due to the integrable nature of this system, which contrasts with the exponential suppression of melting as is in the case of Many-Body Localized time crystals~\cite{driven:expheat:abanin:2015,else:bauer:naik:2016,liu:dtc:2023}.

For each parameter point $(g_{0},\lambda)$ (chosen near putative critical points, deep DTC and spin--liquid--like regions), the pair $(k_{0},\omega)$ is obtained numerically as described in Sect.~\ref{sec:rigidity}. In general, $k_{0}$ is incommensurate with lattice discretization for finite $N$, and so the nearest lattice momentum $k_{0}^R(N)$ is used for time evolution.
\begin{equation}
    k_{0}^{R}(N)=\operatorname*{arg\,min}_{2\pi n/N} |k-k_{0}|.
\end{equation}
The mismatch $\delta k(N)=k_{0} - k_{0}^{R}(N)$ induces a slow dephasing of the subharmonic response nominally locked at $\Omega=\omega/2$, producing a beat envelope and a splitting in the Fourier spectrum. The melting (dephasing) time satisfies $t_m(N)\sim 1/\delta\Omega(N)$, where $\delta\Omega$ is the frequency splitting. For each $N$ (log-spaced from $\mathcal{O}(10^{2})$ to $\mathcal{O}(10^{4})$) we construct $2\times 2$ unitaries $U_{1},U_2$ for $k_{0}^{R}(N)$ with period $T=2\pi/\omega$ and choose $g_{1} = 2\omega$ (equilibrium resonance condition). Starting from the vacuum spinor, we evolve stroboscopically for the $2n_{\text{cycles}}+1$ half-steps ($n_{\text{cycles}}=10^{5}$), storing $C_z(k^R_{0},nT)$. Computations employ GPU acceleration (CuPy~\cite{cupy:learningsys2017}) and Python multiprocessing in $N$. The real-time series is then FFT-transformed. A two-Lorentzian model $L_{1}+L_2$ (implemented by using lmfit~\cite{lmfit:Newville2025}) fits $|{\rm FFT}|$ near $\omega/2$, producing peak centers $\Omega_{1,2}$ and a melting frequency $\delta\Omega = |\Omega_{1} - \Omega_2|$. After filtering ($\delta\Omega> \text{floor}$, finite values), the Random Sample Consensus (RANSAC) regression algorithm of the scikit-learn package~\cite{scikit-learn} is applied to $\log\delta\Omega$ versus $\log N$ to extract a scaling exponent $\alpha$ from the model giving $\delta\Omega \sim N^{\alpha}$.  Figure~\ref{fig:finitesize:ffts} shows the fitted spectra and scaling for several parameter points (panels labeled a-h, x, y), while Fig.~\ref{fig:finitesize:ts} shows the corresponding time series $C_z(k_{0}^R,nT)$ for two lattice sizes. Both figures illustrate the different regimes observed.
\subsection{Observed Regimes}
\begin{figure*}[ht!]
    \centering
    \includegraphics[height=0.75\textheight, keepaspectratio]{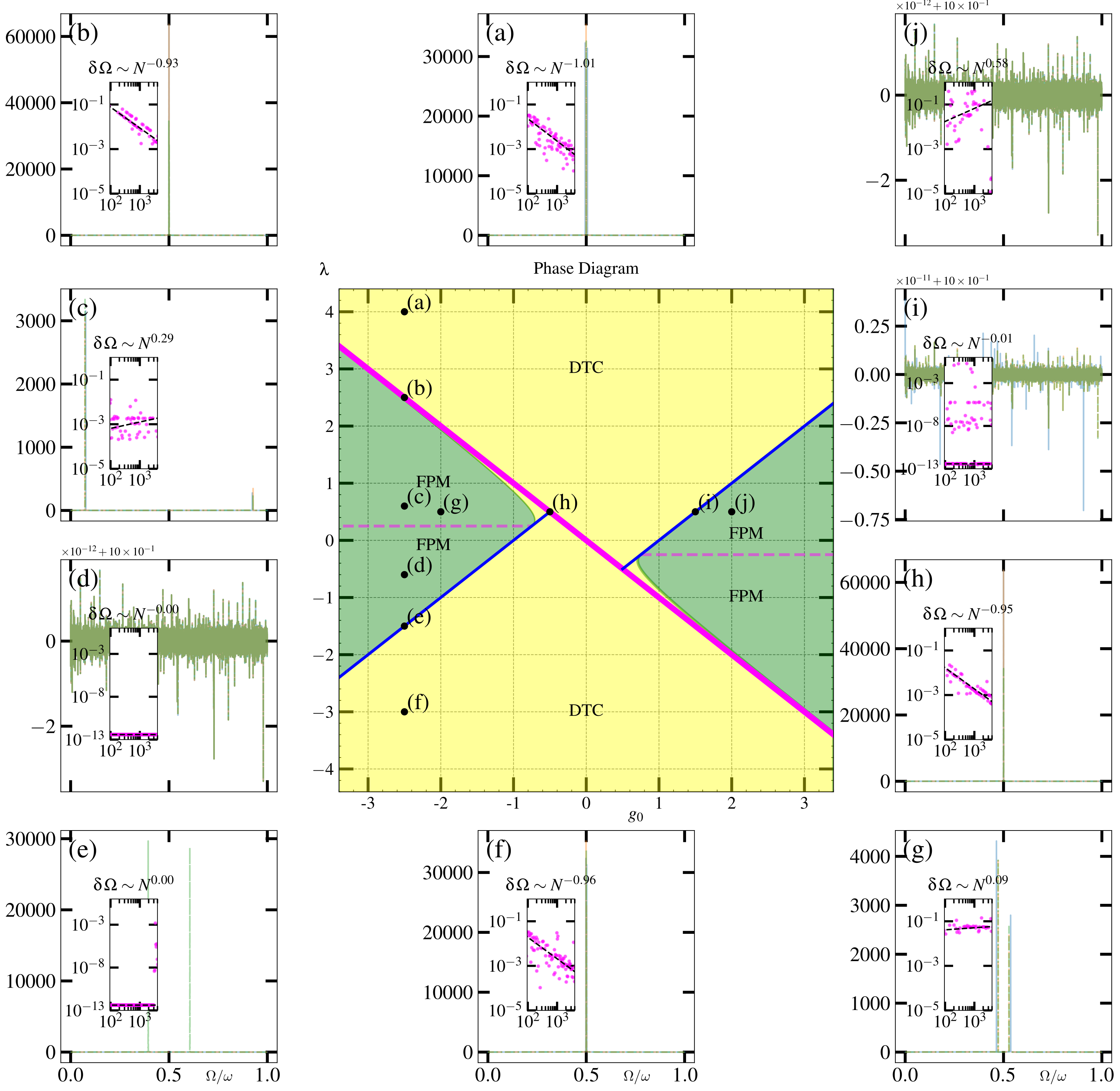}
    \caption{Center: Full floquet phase diagram in the $(g_{0},\lambda)$ plane at equilibrium resonance $g_{1}=2\omega$, indicating regions of Discrete Time-Crystal (DTC, yellow) and Floquet Paramagnet (FPM, green). Solid blue and magenta lines mark analytically obtained Floquet gapless points at high-symmetry and non-HS momenta [Eqs.~(\ref{eq:gapless:hs}, \ref{eq:gapless:nohs})]. Black dots (marked a-j) identify parameter points used in the finite-size analysis. Surrounding panels in counter-clockwise order: normalized power spectra |FFT| of the stroboscopic subharmonic at the optimal momentum (with incommensurate $k_{0}$ replaced by its nearest lattice momentum $k_{0}^{R}(N)$) for several system sizes (solid traces: increasing $N$; dashed verticals: fitted peak centers). Each panel reports the fitted scaling $\delta\Omega\sim N^{\alpha}$ obtained from the two-Lorentzian peak splitting $\delta\Omega=|\Omega_{1}-\Omega_2|$; insets display the log-log data and the RANSAC fit.}
\label{fig:finitesize:ffts}
\end{figure*}
\begin{figure*}[ht!]
    \centering
    \includegraphics[height=0.75\textheight, keepaspectratio]{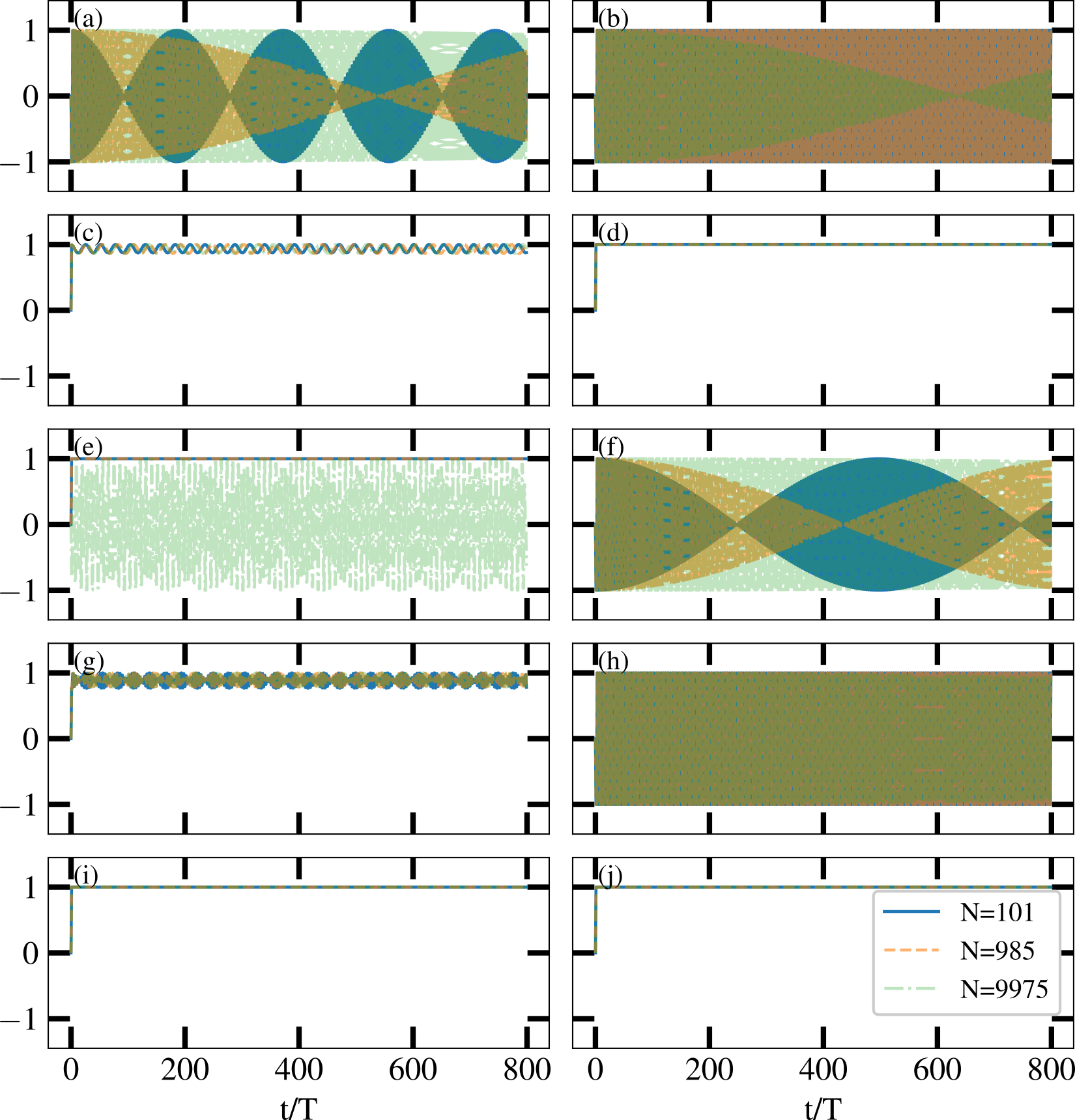}
    \caption{Stroboscopic time series of $C_z\left(k_{0}^{R}, nT\right)$ (defined in Eq.~\eqref{eq:correlations:kspace}) at $k_{0}^R$, the lattice momentum nearest to the optimal $k_{0}$, shown for three system sizes ($N\sim 100$: solid dark-cyan; $N\sim 1000$ dashed dark-orange, and $N\sim 10,000$; dash-dot lime-green). Panels where the parameters are deep in the DTC phase (a,f) display long-lived period-doubled oscillations with a finite-size beat envelope whose period grows with $N$ (data shown up to $n=800$ drive periods). Near-critical panels on the DTC side (b,h) show slower decoherence; and those on the FPM side (e,i) show mostly stationary behavior. The FPM at (c,g) has fast, low-amplitude oscillations, while the FPM at (d,j) is essentially stationary, illustrating how finite-size dephasing discriminates robust DTC scaling from FPM behavior.}
\label{fig:finitesize:ts}
\end{figure*}
The following regimes are revealed from the plots in Figs.~\ref{fig:finitesize:ffts} and~\ref{fig:finitesize:ts}:
\begin{itemize}
\item \textbf{Deep DTC (panels a,f,h):} Clear and persistent two-peak structure in the FFTs of Fig.~\ref{fig:finitesize:ffts}; scaling exponent $\alpha \approx -1$ (algebraic decay). The time series in Fig.~\ref{fig:finitesize:ts} further corroborate this identification, showing long-lived period-doubled oscillations with a finite-size beat envelope whose period grows with $N$. Hence, the suppression of melting is weaker than the exponential suppression observed for time crystals in non-integrable systems~\cite{driven:expheat:abanin:2015,else:bauer:naik:2016,liu:dtc:2023}. We anticipate that incorporating suitable 4-fermion perturbations, in analogy with the spin-orbit coupling introduced in 2D integrable time crystals discussed in~\cite{inttc:chandra:2024}, would enhance this suppression to exponential scaling, which we leave for future work.
\item \textbf{FPM (panels c, d, g, j):} Spectral splitting deteriorates or plateaus, and the extracted $\alpha$ is unstable (RANSAC retains only a small inlier subset). Panel c in Fig.~\ref{fig:finitesize:ts} shows rapid, low-amplitude oscillations with little dependence on $N$ (the oscillatory FPM), while panels d and j show essentially stationary behavior (due to the subharmonic momentum $k_0$ being at an HS point). These behaviors contrast sharply with the DTC regimes, where the beat envelope period increases with $N$.
\item \textbf{On the asymptotic critical line $\lambda\approx -g_0$ (panel b):} Similar negative $\alpha$ on the DTC side.
\item \textbf{On critical lines $\lambda=g_0\pm1$ (panels e,i):} Dependencies are highly sensitive to machine precision , with behavior consistent with FPM appearing at small system sizes and large spectral splits at larger sizes without consistent scaling.
\end{itemize}

\section{Conclusion}
In our study, we have identified a new Discrete Time-Crystal (DTC) phase within a class of one-dimensional quadratic lattice Hamiltonians subjected to periodic driving. The role of integrability is crucial for generating the necessary subharmonics: it enables the decomposition of the Hilbert space into an extensive number of invariant momentum-space subspaces, which in turn facilitates sustained subharmonic response at selected momenta. By analyzing the Floquet quasiparticle spectrum across the parameter space, we demonstrated how NNN interactions open and control quasienergy gaps that pin these modes and suppress dephasing channels, stabilizing the DTC against melting in strictly one-dimensional spin chains. We further identified a Floquet paramagnetic regime and corroborated its nature with finite-size scaling of the subharmonic splitting. The finite-size analysis also revealed algebraic scaling in the suppression of melting, which is weaker than the exponential lifetimes in non-integrable systems such as MBL-stabilized DTCs. However, it should be noted that an analogous algebraic decay occurs in the higher-dimensional integrable systems analyzed in~\cite{inttc:chandra:2024}, where it was transformed into an exponential decay by the perturbative inclusion of non-integrable four-fermion coupling terms. We expect that comparable enhancements will occur in the one-dimensional systems investigated in the present work, a detailed analysis of which is deferred to future research.

These results fall in line with recent theoretical advances in integrable Floquet systems showing that algebraic integrability and engineered couplings can together stabilize time-crystalline order without disorder or prethermal fine-tuning. In particular, the NNN engineering we propose provides a concrete, experimentally accessible route to realize integrability-protected DTCs in one dimension. With the rapid development of programmable quantum simulators and superconducting-qubit platforms capable of tunable NNN couplings and variational control, the stabilization mechanism described here can be tested on Noisy Intermediate-Scale Quantum (NISQ) devices, thereby linking exactly solvable theory to near-term experiments in synthetic quantum matter.

\begin{acknowledgments}
RC and AR acknowledge the use of the BUParamShavak high-performance computing facility at the Department of Physics, The University of Burdwan, Bardhaman, India. SS acknowledges support from the Science and Engineering Research Board (SERB), India, through the Core Research Grant No. CRG / $2021$ / $000996$. MR thanks DST, India, for support through the DST/FFT/NQM/QSM/2024/3 project.
\end{acknowledgments}

\section*{Author Contributions}
\noindent\insertcreditsstatement

\bibliography{references}

@misc{rahaman:2026,
  title        = {Discrete Time Crystal Order in Spin-Chains Enabled by Floquet Flat-Bands},
  author       = {Mahbub Rahaman and Analabha Roy},
  year         = {2026},
  url          = {https://arxiv.org/abs/2603.14307},
  eprint       = {2603.14307},
  archiveprefix = {arXiv},
  primaryclass = {cond-mat.stat-mech}
}

@article{znidaric:2025,
  title        = {Integrability Is Generic in Homogeneous {{U}}(1)-Invariant Nearest-Neighbor Qubit Circuits},
  author       = {{{\v Z} {\v Z}nidari{\v c} {\v c}}, Marko and Duh, Urban and Zadnik, Lenart},
  year         = {2025},
  month        = jul,
  journal      = {Physical Review B},
  publisher    = {American Physical Society},
  volume       = {112},
  number       = {2},
  pages        = {L020302},
  doi          = {10.1103/tqy8-ynpd}
}

@article{anisur:2025,
  title        = {Quasi-Discrete Time Crystals in the Quasiperiodically Driven Lipkin--Meshkov--Glick Model},
  author       = {Anisur, Sk and Liu, Wensheng Vincent and Choudhury, Sayan},
  year         = {2025},
  month        = jun,
  journal      = {Entropy. An International and Interdisciplinary Journal of Entropy and Information Studies},
  publisher    = {Mdpi Ag},
  volume       = {27},
  number       = {6},
  pages        = {609},
  doi          = {10.3390/e27060609},
  issn         = {1099-4300}
}

@misc{Li:2025,
  title        = {Discrete time crystals in one-dimensional classical Floquet systems with nearest-neighbor interactions},
  author       = {Li,  Zhuo-Yi and Zhang,  Yu-Ran},
  year         = {2025},
  publisher    = {arXiv},
  doi          = {10.48550/ARXIV.2505.07524},
  eprint       = {2505.07524},
  archiveprefix = {arXiv},
  primaryclass = {quant-ph}
}

@article{Lu:2025,
  title        = {Implementing Arbitrary Ising Models with a Trapped-Ion Quantum Processor},
  author       = {Lu, Yao and Chen, Wentao and Zhang, Shuaining and Zhang, Kuan and Zhang, Jialiang and Zhang, Jing-Ning and Kim, Kihwan},
  year         = {2025},
  month        = feb,
  journal      = {Phys. Rev. Lett.},
  publisher    = {American Physical Society},
  volume       = {134},
  pages        = {050602},
  doi          = {10.1103/PhysRevLett.134.050602},
  issue        = {5},
  numpages     = {7}
}

@misc{lmfit:newville2025,
  title        = {{{LMFIT}}: {{Non-Linear Least-Squares Minimization}} and {{Curve-Fitting}} for {{Python}}},
  author       = {Newville, Matthew and Otten, Renee and Nelson, Andrew and Stensitzki, Till and Ingargiola, Antonino and Allan, Daniel and Fox, Austin and Carter, Faustin and Rawlik, Michal},
  year         = {2025},
  doi          = {10.5281/zenodo.16175987}
}

@article{pizzi:2025,
  title        = {Genuine Quantum Scars in Many-Body Spin Systems},
  author       = {Pizzi, Andrea and Kwan, Long-Hei and Evrard, Bertrand and Dag, Ceren B. and Knolle, Johannes},
  year         = {2025},
  month        = jul,
  journal      = {Nature Communications},
  publisher    = {Springer Science and Business Media LLC},
  volume       = {16},
  number       = {1},
  pages        = {6722},
  doi          = {10.1038/s41467-025-61765-3},
  issn         = {2041-1723}
}

@article{Weaving:2025,
  title        = {Accurately {S}imulating the {T}ime {E}volution of an {I}sing {M}odel with {E}cho {V}erified {C}lifford {D}ata {R}egression on a {S}uperconducting {Q}uantum {C}omputer},
  author       = {Weaving, Tim and Ralli, Alexis and Love, Peter J. and Succi, Sauro and Coveney, Peter V.},
  year         = {2025},
  month        = may,
  journal      = {{Quantum}},
  publisher    = {{Verein zur F{\"{o}}rderung des Open Access Publizierens in den Quantenwissenschaften}},
  volume       = {9},
  pages        = {1732},
  doi          = {10.22331/q-2025-05-05-1732},
  issn         = {2521-327X}
}

@article{axs:2024,
  title        = {Data-Driven Modeling of Subharmonic Forced Response Due to Nonlinear Resonance},
  author       = {Ax{\aa}s, Joar and B{\"a}uerlein, Bastian and Avila, Kerstin and Haller, George},
  year         = {2024},
  month        = oct,
  journal      = {Scientific Reports},
  publisher    = {Springer Science and Business Media LLC},
  volume       = {14},
  number       = {1},
  pages        = {25991},
  doi          = {10.1038/s41598-024-77639-5},
  issn         = {2045-2322}
}

@article{inttc:chandra:2024,
  title        = {Discrete {{Time Crystal Phase}} of {{Higher Dimensional Integrable Models}}},
  author       = {Chandra, Rahul and Roy, Analabha},
  year         = {2024},
  month        = jul,
  journal      = {Physics Letters A},
  publisher    = {Elsevier BV},
  volume       = {511},
  pages        = {129552},
  doi          = {10.1016/j.physleta.2024.129552},
  issn         = {0375-9601}
}

@misc{qutip5,
  title        = {Qutip 5: {{The Quantum Toolbox}} in {{Python}}},
  author       = {Lambert, Neill and Gigu{\`e}re, Eric and Menczel, Paul and Li, Boxi and Hopf, Patrick and Su{\'a}rez, Gerardo and Gali, Marc and Lishman, Jake and Gadhvi, Rushiraj and Agarwal, Rochisha and Galicia, Asier and Shammah, Nathan and Nation, Paul and Johansson, J. R. and Ahmed, Shahnawaz and Cross, Simon and Pitchford, Alexander and Nori, Franco},
  year         = {2024},
  doi          = {10.48550/arxiv.2412.04705},
  eprint       = {2412.04705},
  primaryclass = {quant-ph},
  archiveprefix = {arXiv}
}

@article{ising:mbeng:2024,
  title        = {The Quantum {{Ising}} Chain for Beginners},
  author       = {Mbeng, Glen Bigan and Russomanno, Angelo and Santoro, Giuseppe E.},
  year         = {2024},
  journal      = {SciPost Phys. Lect. Notes},
  publisher    = {SciPost},
  pages        = {82},
  doi          = {10.21468/SciPostPhysLectNotes.82}
}

@article{rahamanlmg:2024,
  title        = {Phase Crossover Induced by Dynamical Many-Body Localization in Periodically Driven Long-Range Spin Systems},
  author       = {Rahaman, Mahbub and Mori, Takashi and Roy, Analabha},
  year         = {2024},
  month        = mar,
  journal      = {Phys. Rev. B},
  publisher    = {American Physical Society},
  volume       = {109},
  pages        = {104311},
  doi          = {10.1103/PhysRevB.109.104311}
}

@article{rahaman:2024,
  title        = {Time crystal embodies chimeralike state in periodically driven quantum spin system},
  author       = {Rahaman, Mahbub and Sakurai, Akitada and Roy, Analabha},
  year         = {2024},
  month        = jun,
  journal      = {New Journal of Physics},
  publisher    = {IOP Publishing},
  volume       = {26},
  number       = {6},
  pages        = {063035},
  doi          = {10.1088/1367-2630/ad5757}
}

@article{qtckt:vernier:2024,
  title        = {Strong Zero Modes in Integrable Quantum Circuits},
  author       = {Vernier, Eric and Yeh, Hsiu-Chung and Piroli, Lorenzo and Mitra, Aditi},
  year         = {2024},
  month        = aug,
  journal      = {Physical Review Letters},
  publisher    = {American Physical Society},
  volume       = {133},
  number       = {5},
  pages        = {050606},
  doi          = {10.1103/PhysRevLett.133.050606}
}

@article{banerjee:2023,
  title        = {Emergent Conservation in the Floquet Dynamics of Integrable Non-Hermitian Models},
  author       = {Banerjee, Tista and Sengupta, K.},
  year         = {2023},
  month        = apr,
  journal      = {Physical Review B},
  publisher    = {American Physical Society},
  volume       = {107},
  number       = {15},
  pages        = {155117},
  doi          = {10.1103/PhysRevB.107.155117}
}

@article{chandran:2023,
  title        = {Quantum {{Many-Body Scars}}: {{A Quasiparticle Perspective}}},
  author       = {Chandran, Anushya and Iadecola, Thomas and Khemani, Vedika and Moessner, Roderich},
  year         = {2023},
  month        = mar,
  journal      = {Annual Review of Condensed Matter Physics},
  publisher    = {Annual Reviews},
  volume       = {14},
  number       = {1},
  pages        = {443--469},
  doi          = {10.1146/annurev-conmatphys-031620-101617},
  issn         = {1947-5462}
}

@article{deng:2023,
  title        = {Using Models with Static Quantum Many-Body Scars to Generate Time-Crystalline Behavior under Periodic Driving},
  author       = {Deng, Wentai and Yang, Zhi-Cheng},
  year         = {2023},
  month        = nov,
  journal      = {Physical Review B},
  publisher    = {American Physical Society},
  volume       = {108},
  number       = {20},
  pages        = {205129},
  doi          = {10.1103/PhysRevB.108.205129}
}

@article{giachetti:2023,
  title        = {Fractal Nature of High-Order Time Crystal Phases},
  author       = {Giachetti, Guido and Solfanelli, Andrea and Correale, Lorenzo and Defenu, Nicol{\`o}},
  year         = {2023},
  month        = oct,
  journal      = {Physical Review B},
  publisher    = {American Physical Society},
  volume       = {108},
  number       = {14},
  pages        = {L140102},
  doi          = {10.1103/PhysRevB.108.L140102}
}

@article{ha:2023,
  title        = {Many-{{Body Resonances}} in the {{Avalanche}} Instability of Many-Body Localization},
  author       = {Ha, Hyunsoo and Morningstar, Alan and Huse, David A.},
  year         = {2023},
  month        = jun,
  journal      = {Physical Review Letters},
  publisher    = {American Physical Society},
  volume       = {130},
  number       = {25},
  pages        = {250405},
  doi          = {10.1103/PhysRevLett.130.250405}
}

@article{liu:dtc:2023,
  title        = {Discrete Time Crystal Enabled by {{Stark}} Many-Body Localization},
  author       = {Liu, Shuo and Zhang, Shi-Xin and Hsieh, Chang-Yu and Zhang, Shengyu and Yao, Hong},
  year         = {2023},
  month        = mar,
  journal      = {Physical Review Letters},
  volume       = {130},
  number       = {12},
  pages        = {120403},
  doi          = {10.1103/PhysRevLett.130.120403},
  issn         = {0031-9007, 1079-7114},
  eprint       = {2208.02866},
  eprintclass  = {cond-mat, physics:quant-ph}
}

@article{mattes:2023,
  title        = {Entangled Time-Crystal Phase in an Open Quantum Light-Matter System},
  author       = {Mattes, Robert and Lesanovsky, Igor and Carollo, Federico},
  year         = {2023},
  month        = dec,
  journal      = {Physical Review A: Atomic, Molecular, and Optical Physics},
  publisher    = {American Physical Society},
  volume       = {108},
  number       = {6},
  pages        = {062216},
  doi          = {10.1103/PhysRevA.108.062216}
}

@article{muller:2023,
  title        = {Exact Conservation Laws for Neural Network Integrators of Dynamical Systems},
  author       = {M{\"u}ller, Eike Hermann},
  year         = {2023},
  month        = sep,
  journal      = {Journal of Computational Physics},
  publisher    = {Elsevier BV},
  volume       = {488},
  pages        = {112234},
  doi          = {10.1016/j.jcp.2023.112234},
  issn         = {0021-9991}
}

@article{stasiuk:2023,
  title        = {Observation of a Prethermal {{u(1)}} Discrete Time Crystal},
  author       = {Stasiuk, Andrew and Cappellaro, Paola},
  year         = {2023},
  month        = oct,
  journal      = {Physical Review X},
  publisher    = {American Physical Society},
  volume       = {13},
  number       = {4},
  pages        = {041016},
  doi          = {10.1103/PhysRevX.13.041016}
}

@article{vu:2023,
  title        = {Dissipative Prethermal Discrete Time Crystal},
  author       = {Vu, DinhDuy and Das Sarma, Sankar},
  year         = {2023},
  month        = mar,
  journal      = {Physical Review Letters},
  publisher    = {American Physical Society},
  volume       = {130},
  number       = {13},
  pages        = {130401},
  doi          = {10.1103/PhysRevLett.130.130401}
}

@article{tc:rmp:colloquium:2023,
  title        = {Colloquium: {{Quantum}} and Classical Discrete Time Crystals},
  author       = {Zaletel, Michael P. and Lukin, Mikhail and Monroe, Christopher and Nayak, Chetan and Wilczek, Frank and Yao, Norman Y.},
  year         = {2023},
  month        = jul,
  journal      = {Reviews of Modern Physics},
  publisher    = {American Physical Society},
  volume       = {95},
  number       = {3},
  pages        = {031001},
  doi          = {10.1103/RevModPhys.95.031001}
}

@article{trotter:zhao:2023,
  title        = {Making Trotterization Adaptive and Energy-Self-Correcting for Nisq Devices and Beyond},
  author       = {Zhao, Hongzheng and Bukov, Marin and Heyl, Markus and Moessner, Roderich},
  year         = {2023},
  month        = aug,
  journal      = {PRX Quantum},
  publisher    = {American Physical Society},
  volume       = {4},
  number       = {3},
  pages        = {030319},
  doi          = {10.1103/PRXQuantum.4.030319}
}

@article{birnkammer:2022,
  title        = {Prethermalization in One-Dimensional Quantum Many-Body Systems with Confinement},
  author       = {Birnkammer, Stefan and Bastianello, Alvise and Knap, Michael},
  year         = {2022},
  month        = dec,
  journal      = {Nature Communications},
  publisher    = {Springer Science and Business Media LLC},
  volume       = {13},
  number       = {1},
  pages        = {7663},
  doi          = {10.1038/s41467-022-35301-6},
  issn         = {2041-1723}
}

@misc{numdifftools,
  title        = {Numdifftools},
  author       = {Brodtkorb, Per A. and D'Errico, John},
  year         = {2022},
  accessed     = {2025-09-28},
  howpublished = {\url{https://github.com/pbrod/numdifftools}}
}

@article{bull2022,
  title        = {Tuning between Continuous Time Crystals and Many-Body Scars in Long-Range {{xyz}} Spin Chains},
  author       = {Bull, Kieran and Hallam, Andrew and {Papi{\'c} {\'c}}, Zlatko and Martin, Ivar},
  year         = {2022},
  month        = sep,
  journal      = {Physical Review Letters},
  publisher    = {American Physical Society},
  volume       = {129},
  number       = {14},
  pages        = {140602},
  doi          = {10.1103/PhysRevLett.129.140602}
}

@article{frey:2022,
  title        = {Realization of a Discrete Time Crystal on 57 Qubits of a Quantum Computer},
  author       = {Frey, Philipp and Rachel, Stephan},
  year         = {2022},
  journal      = {Science Advances},
  volume       = {8},
  number       = {9},
  pages        = {eabm7652},
  doi          = {10.1126/sciadv.abm7652}
}

@article{haldar2022,
  title        = {Statistical Mechanics of Floquet Quantum Matter: {{Exact}} and Emergent Conservation Laws},
  author       = {Haldar, Asmi and Das, Arnab},
  year         = {2022},
  month        = apr,
  journal      = {Journal of Physics: Condensed Matter},
  publisher    = {IOP Publishing},
  volume       = {34},
  number       = {23},
  pages        = {234001},
  doi          = {10.1088/1361-648x/ac03d2},
  issn         = {1361-648x}
}

@article{krajewski:2022,
  title        = {Restoring Ergodicity in a Strongly Disordered Interacting Chain},
  author       = {Krajewski, B. and Vidmar, L. and {Bon{\v c} {\v c}a}, J. and Mierzejewski, M.},
  year         = {2022},
  month        = dec,
  journal      = {Physical Review Letters},
  publisher    = {American Physical Society},
  volume       = {129},
  number       = {26},
  pages        = {260601},
  doi          = {10.1103/PhysRevLett.129.260601}
}

@article{MuozArias:2022,
  title        = {Floquet time crystals in driven spin systems with all-to-all $p-$body interactions},
  author       = {Mu\~noz-Arias,  Manuel H. and Chinni,  Karthik and Poggi,  Pablo M.},
  year         = {2022},
  month        = apr,
  journal      = {Physical Review Research},
  publisher    = {American Physical Society (APS)},
  volume       = {4},
  number       = {2},
  doi          = {10.1103/physrevresearch.4.023018},
  issn         = {2643-1564}
}

@article{trotter:pastori:2022,
  title        = {Characterization and Verification of Trotterized Digital Quantum Simulation via Hamiltonian and Liouvillian Learning},
  author       = {Pastori, Lorenzo and Olsacher, Tobias and Kokail, Christian and Zoller, Peter},
  year         = {2022},
  month        = aug,
  journal      = {PRX Quantum},
  publisher    = {American Physical Society},
  volume       = {3},
  number       = {3},
  pages        = {030324},
  doi          = {10.1103/PRXQuantum.3.030324}
}

@article{santini2022,
  title        = {Clean Two-Dimensional Floquet Time Crystal},
  author       = {Santini, Alessandro and Santoro, Giuseppe E. and Collura, Mario},
  year         = {2022},
  month        = oct,
  journal      = {Physical Review B},
  publisher    = {American Physical Society},
  volume       = {106},
  number       = {13},
  pages        = {134301},
  doi          = {10.1103/PhysRevB.106.134301}
}

@article{yates:2022,
  title        = {Long-Lived Period-Doubled Edge Modes of Interacting and Disorder-Free Floquet Spin Chains},
  author       = {Yates, Daniel J. and Abanov, Alexander G. and Mitra, Aditi},
  year         = {2022},
  month        = feb,
  journal      = {Communications Physics},
  publisher    = {Springer Science and Business Media LLC},
  volume       = {5},
  number       = {1},
  pages        = {43},
  doi          = {10.1038/s42005-022-00818-1},
  issn         = {2399-3650}
}

@article{cerezo:2021,
  title        = {Variational quantum algorithms},
  author       = {Cerezo,  M. and Arrasmith,  Andrew and Babbush,  Ryan and Benjamin,  Simon C. and Endo,  Suguru and Fujii,  Keisuke and McClean,  Jarrod R. and Mitarai,  Kosuke and Yuan,  Xiao and Cincio,  Lukasz and Coles,  Patrick J.},
  year         = {2021},
  month        = aug,
  journal      = {Nature Reviews Physics},
  publisher    = {Springer Science and Business Media LLC},
  volume       = {3},
  number       = {9},
  pages        = {625–644},
  doi          = {10.1038/s42254-021-00348-9},
  issn         = {2522-5820}
}

@article{kebler:2021,
  title        = {Observation of a Dissipative Time Crystal},
  author       = {Ke{\ss}ler, Hans and Kongkhambut, Phatthamon and Georges, Christoph and Mathey, Ludwig and Cosme, Jayson G. and Hemmerich, Andreas},
  year         = {2021},
  month        = jul,
  journal      = {Physical Review Letters},
  publisher    = {American Physical Society},
  volume       = {127},
  number       = {4},
  pages        = {043602},
  doi          = {10.1103/PhysRevLett.127.043602}
}

@article{kumar:2021,
  title        = {Multi-Critical Topological Transition at Quantum Criticality},
  author       = {Kumar, Ranjith R. and Kartik, Y. R. and Rahul, S. and Sarkar, Sujit},
  year         = {2021},
  month        = jan,
  journal      = {Scientific Reports},
  publisher    = {Springer Science and Business Media LLC},
  volume       = {11},
  number       = {1},
  pages        = {1004},
  doi          = {10.1038/s41598-020-80337-7},
  issn         = {2045-2322}
}

@article{kyprianidis:2021,
  title        = {Observation of a Prethermal Discrete Time Crystal},
  author       = {Kyprianidis, A. and Machado, F. and Morong, W. and Becker, P. and Collins, K. S. and Else, D. V. and Feng, L. and Hess, P. W. and Nayak, C. and Pagano, G. and Yao, N. Y. and Monroe, C.},
  year         = {2021},
  journal      = {Science},
  volume       = {372},
  number       = {6547},
  pages        = {1192--1196},
  doi          = {10.1126/science.abg8102}
}

@article{maskara:2021,
  title        = {Discrete Time-Crystalline Order Enabled by Quantum Many-Body Scars: {{Entanglement}} Steering via Periodic Driving},
  author       = {Maskara, N. and Michailidis, A. A. and Ho, W. W. and Bluvstein, D. and Choi, S. and Lukin, M. D. and Serbyn, M.},
  year         = {2021},
  month        = aug,
  journal      = {Physical Review Letters},
  publisher    = {American Physical Society},
  volume       = {127},
  number       = {9},
  pages        = {090602},
  doi          = {10.1103/PhysRevLett.127.090602}
}

@article{OjedaCollado:2021,
  title        = {Emergent parametric resonances and time-crystal phases in driven Bardeen-Cooper-Schrieffer systems},
  author       = {Ojeda Collado,  H. P. and Usaj,  Gonzalo and Balseiro,  C. A. and Zanette,  Damián H. and Lorenzana,  José},
  year         = {2021},
  month        = nov,
  journal      = {Physical Review Research},
  publisher    = {American Physical Society (APS)},
  volume       = {3},
  number       = {4},
  doi          = {10.1103/physrevresearch.3.l042023},
  issn         = {2643-1564}
}

@article{pizzi:2021,
  title        = {Higher-Order and Fractional Discrete Time Crystals in Clean Long-Range Interacting Systems},
  author       = {Pizzi, Andrea and Knolle, Johannes and Nunnenkamp, Andreas},
  year         = {2021},
  month        = apr,
  journal      = {Nature Communications},
  publisher    = {Springer Science and Business Media LLC},
  volume       = {12},
  number       = {1},
  pages        = {2341},
  doi          = {10.1038/s41467-021-22583-5},
  issn         = {2041-1723}
}

@article{prethermal:pizzi:2021,
  title        = {Classical Prethermal Phases of Matter},
  author       = {Pizzi, Andrea and Nunnenkamp, Andreas and Knolle, Johannes},
  year         = {2021},
  month        = sep,
  journal      = {Physical Review Letters},
  publisher    = {American Physical Society},
  volume       = {127},
  number       = {14},
  pages        = {140602},
  doi          = {10.1103/PhysRevLett.127.140602}
}

@article{randall:2021,
  title        = {Many-Body--Localized Discrete Time Crystal with a Programmable Spin-Based Quantum Simulator},
  author       = {Randall, J. and Bradley, C. E. and {van der Gronden}, F. V. and Galicia, A. and Abobeih, M. H. and Markham, M. and Twitchen, D. J. and Machado, F. and Yao, N. Y. and Taminiau, T. H.},
  year         = {2021},
  journal      = {Science},
  volume       = {374},
  number       = {6574},
  pages        = {1474--1478},
  doi          = {10.1126/science.abk0603}
}

@article{wang:2021,
  title        = {Many-Body Effects and Quantum Fluctuations for Discrete Time Crystals in Bose--Einstein Condensates},
  author       = {Wang, Jia and Hannaford, Peter and Dalton, Bryan J},
  year         = {2021},
  month        = jun,
  journal      = {New Journal of Physics},
  publisher    = {IOP Publishing},
  volume       = {23},
  number       = {6},
  pages        = {063012},
  doi          = {10.1088/1367-2630/abea45},
  issn         = {1367-2630}
}

@article{dtcbec:wang:2021,
  title        = {Discrete Time Crystals in Bose-Einstein Condensates and the Symmetry-Breaking Edge in a Simple Two-Mode Theory},
  author       = {Wang, Jia and Sacha, Krzysztof and Hannaford, Peter and Dalton, Bryan J.},
  year         = {2021},
  month        = nov,
  journal      = {Physical Review A: Atomic, Molecular, and Optical Physics},
  publisher    = {American Physical Society},
  volume       = {104},
  number       = {5},
  pages        = {053327},
  doi          = {10.1103/PhysRevA.104.053327}
}

@article{floquet:yates:2021,
  title        = {Strong and Almost Strong Modes of Floquet Spin Chains in Krylov Subspaces},
  author       = {Yates, Daniel J. and Mitra, Aditi},
  year         = {2021},
  month        = nov,
  journal      = {Physical Review B},
  publisher    = {American Physical Society},
  volume       = {104},
  number       = {19},
  pages        = {195121},
  doi          = {10.1103/PhysRevB.104.195121}
}

@article{else:2020,
  title        = {Discrete Time Crystals},
  author       = {Else, Dominic V. and Monroe, Christopher and Nayak, Chetan and Yao, Norman Y.},
  year         = {2020},
  month        = mar,
  journal      = {Annual Review of Condensed Matter Physics},
  publisher    = {Annual Reviews},
  volume       = {11},
  number       = {1},
  pages        = {467--499},
  doi          = {10.1146/annurev-conmatphys-031119-050658},
  issn         = {1947-5462}
}

@article{huetado:2020,
  title        = {Building Continuous Time Crystals from Rare Events},
  author       = {{Hurtado-Guti{\'e}rrez}, R. and Carollo, F. and {P{\'e}rez-Espigares}, C. and Hurtado, P. I.},
  year         = {2020},
  month        = oct,
  journal      = {Physical Review Letters},
  publisher    = {American Physical Society},
  volume       = {125},
  number       = {16},
  pages        = {160601},
  doi          = {10.1103/PhysRevLett.125.160601}
}

@article{kshetrimayum:2020,
  title        = {Stark Time Crystals: {{Symmetry}} Breaking in Space and Time},
  author       = {Kshetrimayum, A. and Eisert, J. and Kennes, D. M.},
  year         = {2020},
  month        = nov,
  journal      = {Physical Review B},
  publisher    = {American Physical Society (APS)},
  volume       = {102},
  number       = {19},
  pages        = {195116},
  doi          = {10.1103/physrevb.102.195116},
  issn         = {2469-9969}
}

@article{luitz:2020,
  title        = {Prethermalization without Temperature},
  author       = {Luitz, David J. and Moessner, Roderich and Sondhi, S. L. and Khemani, Vedika},
  year         = {2020},
  month        = may,
  journal      = {Physical Review X},
  publisher    = {American Physical Society},
  volume       = {10},
  number       = {2},
  pages        = {021046},
  doi          = {10.1103/PhysRevX.10.021046}
}

@article{machado:2020,
  title        = {Long-Range Prethermal Phases of Nonequilibrium Matter},
  author       = {Machado, Francisco and Else, Dominic V. and {Kahanamoku-Meyer}, Gregory D. and Nayak, Chetan and Yao, Norman Y.},
  year         = {2020},
  month        = feb,
  journal      = {Physical Review X},
  publisher    = {American Physical Society},
  volume       = {10},
  number       = {1},
  pages        = {011043},
  doi          = {10.1103/PhysRevX.10.011043}
}

@article{pizzi:2020,
  title        = {Time Crystallinity and Finite-Size Effects in Clean Floquet Systems},
  author       = {Pizzi, Andrea and Malz, Daniel and De Tomasi, Giuseppe and Knolle, Johannes and Nunnenkamp, Andreas},
  year         = {2020},
  month        = dec,
  journal      = {Physical Review B},
  publisher    = {American Physical Society},
  volume       = {102},
  number       = {21},
  pages        = {214207},
  doi          = {10.1103/PhysRevB.102.214207}
}

@article{shtanko:2020,
  title        = {Unitary Subharmonic Response and Floquet Majorana Modes},
  author       = {Shtanko, Oles and Movassagh, Ramis},
  year         = {2020},
  month        = aug,
  journal      = {Physical Review Letters},
  publisher    = {American Physical Society},
  volume       = {125},
  number       = {8},
  pages        = {086804},
  doi          = {10.1103/PhysRevLett.125.086804}
}

@article{scipy:nmeth:2022,
  title        = {{{SciPy}} 1.0: {{Fundamental}} Algorithms for Scientific Computing in Python},
  author       = {Virtanen, Pauli and Gommers, Ralf and Oliphant, Travis E. and Haberland, Matt and Reddy, Tyler and Cournapeau, David and Burovski, Evgeni and Peterson, Pearu and Weckesser, Warren and Bright, Jonathan and {van der Walt}, St{\'e}fan J. and Brett, Matthew and Wilson, Joshua and Millman, K. Jarrod and Mayorov, Nikolay and Nelson, Andrew R. J. and Jones, Eric and Kern, Robert and Larson, Eric and Carey, C J and Polat, {\.I}lhan and Feng, Yu and Moore, Eric W. and VanderPlas, Jake and Laxalde, Denis and Perktold, Josef and Cimrman, Robert and Henriksen, Ian and Quintero, E. A. and Harris, Charles R. and Archibald, Anne M. and Ribeiro, Ant{\^o}nio H. and Pedregosa, Fabian and {van Mulbregt}, Paul and {SciPy 1.0 Contributors}},
  year         = {2020},
  journal      = {Nature Methods},
  volume       = {17},
  pages        = {261--272},
  doi          = {10.1038/s41592-019-0686-2}
}

@article{denardis:2019,
  title        = {Diffusion in Generalized Hydrodynamics and Quasiparticle Scattering},
  author       = {De Nardis, Jacopo and Bernard, Denis and Doyon, Benjamin},
  year         = {2019},
  month        = apr,
  journal      = {SciPost Physics},
  publisher    = {SciPost},
  volume       = {6},
  number       = {4},
  pages        = {049},
  doi          = {10.21468/scipostphys.6.4.049},
  issn         = {2542-4653}
}

@article{kebler:2019,
  title        = {Emergent Limit Cycles and Time Crystal Dynamics in an Atom-Cavity System},
  author       = {Ke{\ss}ler, Hans and Cosme, Jayson G. and Hemmerling, Michal and Mathey, Ludwig and Hemmerich, Andreas},
  year         = {2019},
  month        = may,
  journal      = {Physical Review A: Atomic, Molecular, and Optical Physics},
  publisher    = {American Physical Society},
  volume       = {99},
  number       = {5},
  pages        = {053605},
  doi          = {10.1103/PhysRevA.99.053605}
}

@incollection{kockum:2019,
  title        = {Quantum Bits with Josephson Junctions},
  author       = {Kockum, Anton Frisk and Nori, Franco},
  year         = {2019},
  booktitle    = {Fundamentals and Frontiers of the Josephson Effect},
  publisher    = {Springer International Publishing},
  pages        = {703--741},
  doi          = {10.1007/978-3-030-20726-7_17},
  issn         = {2196-2812},
  note         = {{ISBN:} 9783030207267}
}

@article{takagi:2019,
  title        = {Concept and Realization of Kitaev Quantum Spin Liquids},
  author       = {Takagi, Hidenori and Takayama, Tomohiro and Jackeli, George and Khaliullin, Giniyat and Nagler, Stephen E.},
  year         = {2019},
  month        = mar,
  journal      = {Nature Reviews Physics},
  publisher    = {Springer Science and Business Media LLC},
  volume       = {1},
  number       = {4},
  pages        = {264--280},
  doi          = {10.1038/s42254-019-0038-2},
  issn         = {2522-5820}
}

@article{giergiel:2018,
  title        = {Time Crystal Platform: {{From}} Quasicrystal Structures in Time to Systems with Exotic Interactions},
  author       = {Giergiel, Krzysztof and Miroszewski, Artur and Sacha, Krzysztof},
  year         = {2018},
  month        = apr,
  journal      = {Physical Review Letters},
  publisher    = {American Physical Society},
  volume       = {120},
  number       = {14},
  pages        = {140401},
  doi          = {10.1103/PhysRevLett.120.140401}
}

@article{hermanns:2018,
  title        = {Physics of the Kitaev Model: {{Fractionalization}}, Dynamic Correlations, and Material Connections},
  author       = {Hermanns, M. and Kimchi, I. and Knolle, J.},
  year         = {2018},
  month        = mar,
  journal      = {Annual Review of Condensed Matter Physics},
  publisher    = {Annual Reviews},
  volume       = {9},
  number       = {1},
  pages        = {17--33},
  doi          = {10.1146/annurev-conmatphys-033117-053934},
  issn         = {1947-5462}
}

@article{sarkar:2018,
  title        = {Quantization of Geometric Phase with Integer and Fractional Topological Characterization in a Quantum Ising Chain with Long-Range Interaction},
  author       = {Sarkar, Sujit},
  year         = {2018},
  month        = apr,
  journal      = {Scientific Reports},
  publisher    = {Springer Science and Business Media LLC},
  volume       = {8},
  number       = {1},
  pages        = {5864},
  doi          = {10.1038/s41598-018-24136-1},
  issn         = {2045-2322}
}

@article{ostman:2018,
  title        = {Ising-like Behaviour of Mesoscopic Magnetic Chains},
  author       = {{\"O}stman, Erik and Arnalds, Unnar B and Kapaklis, Vassilios and Taroni, Andrea and Hj{\"o}rvarsson, Bj{\"o}rgvin},
  year         = {2018},
  month        = aug,
  journal      = {Journal of Physics: Condensed Matter},
  publisher    = {IOP Publishing},
  volume       = {30},
  number       = {36},
  pages        = {365301},
  doi          = {10.1088/1361-648X/aad0c1}
}

@article{choi:2017,
  title        = {Observation of Discrete Time-Crystalline Order in a Disordered Dipolar Many-Body System},
  author       = {Choi, Soonwon and Choi, Joonhee and Landig, Renate and Kucsko, Georg and Zhou, Hengyun and Isoya, Junichi and Jelezko, Fedor and Onoda, Shinobu and Sumiya, Hitoshi and Khemani, Vedika and {von Keyserlingk}, Curt and Yao, Norman Y. and Demler, Eugene and Lukin, Mikhail D.},
  year         = {2017},
  month        = mar,
  journal      = {Nature},
  publisher    = {Springer Science and Business Media LLC},
  volume       = {543},
  number       = {7644},
  pages        = {221--225},
  doi          = {10.1038/nature21426},
  issn         = {1476-4687}
}

@article{integrable:floquet:gritsev:2017,
  title        = {Integrable Floquet Dynamics},
  author       = {Gritsev, Vladimir and Polkovnikov, Anatoli},
  year         = {2017},
  journal      = {SciPost Physics},
  publisher    = {SciPost},
  volume       = {2},
  pages        = {021},
  doi          = {10.21468/SciPostPhys.2.3.021}
}

@article{nguyen:2017,
  title        = {Competing Interactions in Artificial Spin Chains},
  author       = {Nguyen, V.-D. and Perrin, Y. and Le Denmat, S. and Canals, B. and Rougemaille, N.},
  year         = {2017},
  month        = jul,
  journal      = {Physical Review B},
  publisher    = {American Physical Society},
  volume       = {96},
  number       = {1},
  pages        = {014402},
  doi          = {10.1103/PhysRevB.96.014402}
}

@inproceedings{cupy:learningsys2017,
  title        = {{{CuPy}}: {{A NumPy-Compatible Library}} for {{NVIDIA GPU Calculations}}},
  author       = {Okuta, Ryosuke and Unno, Yuya and Nishino, Daisuke and Hido, Shohei and Loomis, Crissman},
  year         = {2017},
  booktitle    = {LearningSys Workshop, NeurIPS},
  note         = {Proceedings of Workshop on Machine Learning Systems (LearningSys) in the Thirty-First Annual Conference on Neural Information Processing Systems (NIPS). \url{http://learningsys.org/nips17/assets/papers/paper_16.pdf}}
}

@article{Russomanno:2017,
  title        = {Floquet time crystal in the Lipkin-Meshkov-Glick model},
  author       = {Russomanno,  Angelo and Iemini,  Fernando and Dalmonte,  Marcello and Fazio,  Rosario},
  year         = {2017},
  month        = jun,
  journal      = {Physical Review B},
  publisher    = {American Physical Society (APS)},
  volume       = {95},
  number       = {21},
  doi          = {10.1103/physrevb.95.214307},
  issn         = {2469-9969}
}

@article{yao:potter:potirniche:vishwanath:2017,
  title        = {Discrete {{Time Crystals}}: {{Rigidity}}, {{Criticality}}, and {{Realizations}}},
  author       = {Yao, N. Y. and Potter, A. C. and Potirniche, I.-D. and Vishwanath, A.},
  year         = {2017},
  month        = jan,
  journal      = {Physical Review Letters},
  publisher    = {American Physical Society},
  volume       = {118},
  number       = {3},
  pages        = {030401},
  doi          = {10.1103/PhysRevLett.118.030401}
}

@article{zhang:2017,
  title        = {Observation of a Discrete Time Crystal},
  author       = {Zhang, J. and Hess, P. W. and Kyprianidis, A. and Becker, P. and Lee, A. and Smith, J. and Pagano, G. and Potirniche, I.-D. and Potter, A. C. and Vishwanath, A. and Yao, N. Y. and Monroe, C.},
  year         = {2017},
  month        = mar,
  journal      = {Nature},
  publisher    = {Springer Science and Business Media LLC},
  volume       = {543},
  number       = {7644},
  pages        = {217--220},
  doi          = {10.1038/nature21413},
  issn         = {1476-4687}
}

@article{else:bauer:naik:2016,
  title        = {Floquet {{Time Crystals}}},
  author       = {Else, Dominic V. and Bauer, Bela and Nayak, Chetan},
  year         = {2016},
  month        = aug,
  journal      = {Physical Review Letters},
  publisher    = {American Physical Society},
  volume       = {117},
  number       = {9},
  pages        = {090402},
  doi          = {10.1103/PhysRevLett.117.090402}
}

@article{ott:2016,
  title        = {Single Atom Detection in Ultracold Quantum Gases: A Review of Current Progress},
  author       = {Ott, Herwig},
  year         = {2016},
  month        = apr,
  journal      = {Reports on Progress in Physics},
  publisher    = {IOP Publishing},
  volume       = {79},
  number       = {5},
  pages        = {054401},
  doi          = {10.1088/0034-4885/79/5/054401},
  issn         = {1361-6633}
}

@book{zhu:2016,
  title        = {Bogoliubov-de Gennes Method and Its Applications},
  author       = {Zhu, Jian-Xin},
  year         = {2016},
  publisher    = {Springer International Publishing},
  doi          = {10.1007/978-3-319-31314-6},
  issn         = {1616-6361},
  note         = {{ISBN:} 9783319313146}
}

@article{driven:expheat:abanin:2015,
  title        = {Exponentially Slow Heating in Periodically Driven Many-Body Systems},
  author       = {Abanin, Dmitry A. and De Roeck, Wojciech and {\c c}ois Huveneers, Fran{\c c}},
  year         = {2015},
  month        = dec,
  journal      = {Physical Review Letters},
  publisher    = {American Physical Society},
  volume       = {115},
  number       = {25},
  pages        = {256803},
  doi          = {10.1103/PhysRevLett.115.256803}
}

@article{bukov:2015,
  title        = {Universal High-Frequency Behavior of Periodically Driven Systems: {{From}} Dynamical Stabilization to Floquet Engineering},
  author       = {Bukov, Marin and D'Alessio, Luca and Polkovnikov, Anatoli},
  year         = {2015},
  journal      = {Advances in Physics},
  publisher    = {Taylor \& Francis},
  volume       = {64},
  number       = {2},
  pages        = {139--226},
  doi          = {10.1080/00018732.2015.1055918}
}

@article{nandkishore:2015,
  title        = {Many-{{Body Localization}} and {{Thermalization}} in {{Quantum Statistical Mechanics}}},
  author       = {Nandkishore, Rahul and Huse, David A.},
  year         = {2015},
  month        = mar,
  journal      = {Annual Review of Condensed Matter Physics},
  publisher    = {Annual Reviews},
  volume       = {6},
  number       = {1},
  pages        = {15--38},
  doi          = {10.1146/annurev-conmatphys-031214-014726},
  issn         = {1947-5462}
}

@article{tc:watanabe:oshikawa:2015,
  title        = {Absence of Quantum Time Crystals},
  author       = {Watanabe, Haruki and Oshikawa, Masaki},
  year         = {2015},
  month        = jun,
  journal      = {Physical Review Letters},
  publisher    = {American Physical Society},
  volume       = {114},
  number       = {25},
  pages        = {251603},
  doi          = {10.1103/PhysRevLett.114.251603}
}

@article{georgescu:2014,
  title        = {Quantum Simulation},
  author       = {Georgescu, I.M. and Ashhab, S. and Nori, Franco},
  year         = {2014},
  month        = mar,
  journal      = {Reviews of Modern Physics},
  publisher    = {American Physical Society (APS)},
  volume       = {86},
  number       = {1},
  pages        = {153--185},
  doi          = {10.1103/revmodphys.86.153},
  issn         = {1539-0756}
}

@article{ultracold:bissbort:2013,
  title        = {Emulating Solid-State Physics with a Hybrid System of Ultracold Ions and Atoms},
  author       = {Bissbort, U. and Cocks, D. and Negretti, A. and Idziaszek, Z. and Calarco, T. and {Schmidt-Kaler}, F. and Hofstetter, W. and Gerritsma, R.},
  year         = {2013},
  month        = aug,
  journal      = {Physical Review Letters},
  publisher    = {American Physical Society (APS)},
  volume       = {111},
  number       = {8},
  pages        = {080501},
  doi          = {10.1103/physrevlett.111.080501},
  issn         = {1079-7114}
}

@article{tc:patrick:comment:2013,
  title        = {Comment on ``{{Quantum Time Crystals}}''},
  author       = {Bruno, Patrick},
  year         = {2013},
  month        = mar,
  journal      = {Physical Review Letters},
  publisher    = {American Physical Society},
  volume       = {110},
  number       = {11},
  pages        = {118901},
  doi          = {10.1103/PhysRevLett.110.118901}
}

@article{roy:2013,
  title        = {Periodic dynamics of fermionic superfluids in the BCS regime},
  author       = {Roy, A and Dasgupta, R and Modak, S and Das, A and Sengupta, K},
  year         = {2013},
  month        = apr,
  journal      = {Journal of Physics: Condensed Matter},
  publisher    = {IOP Publishing},
  volume       = {25},
  number       = {20},
  pages        = {205703},
  doi          = {10.1088/0953-8984/25/20/205703},
  issn         = {1361-648X}
}

@article{blatt:2012,
  title        = {Quantum Simulations with Trapped Ions},
  author       = {Blatt, R. and Roos, C. F.},
  year         = {2012},
  month        = apr,
  journal      = {Nature Physics},
  publisher    = {Springer Science and Business Media LLC},
  volume       = {8},
  number       = {4},
  pages        = {277--284},
  doi          = {10.1038/nphys2252},
  issn         = {1745-2481}
}

@article{3body:niu:2012,
  title        = {Majorana Zero Modes in a Quantum {{Ising}} Chain with Longer-Ranged Interactions},
  author       = {Niu, Yuezhen and Chung, Suk Bum and Hsu, Chen-Hsuan and Mandal, Ipsita and Raghu, S. and Chakravarty, Sudip},
  year         = {2012},
  month        = jan,
  journal      = {Physical Review B},
  publisher    = {American Physical Society},
  volume       = {85},
  number       = {3},
  pages        = {035110},
  doi          = {10.1103/PhysRevB.85.035110}
}

@incollection{pachos:2012,
  title        = {Kitaev's Honeycomb Lattice Model},
  author       = {Pachos, Jiannis K.},
  year         = {2012},
  booktitle    = {Introduction to Topological Quantum Computation},
  publisher    = {Cambridge University Press},
  address      = {Cambridge},
  pages        = {102--128},
  doi          = {10.1017/CBO9780511792908},
  note         = {{ISBN:} 9780511792908}
}

@article{tc:shapere:2012,
  title        = {Classical {{Time Crystals}}},
  author       = {Shapere, Alfred and Wilczek, Frank},
  year         = {2012},
  month        = oct,
  journal      = {Physical Review Letters},
  publisher    = {American Physical Society},
  volume       = {109},
  number       = {16},
  pages        = {160402},
  doi          = {10.1103/PhysRevLett.109.160402}
}

@article{tc:wilczek:2012,
  title        = {Quantum {{Time Crystals}}},
  author       = {Wilczek, Frank},
  year         = {2012},
  month        = oct,
  journal      = {Physical Review Letters},
  publisher    = {American Physical Society},
  volume       = {109},
  number       = {16},
  pages        = {160401},
  doi          = {10.1103/PhysRevLett.109.160401}
}

@article{scikit-learn,
  title        = {Scikit-{{Learn}}: {{Machine Learning}} in {{Python}}},
  author       = {Pedregosa, F. and Varoquaux, G. and Gramfort, A. and Michel, V. and Thirion, B. and Grisel, O. and Blondel, M. and Prettenhofer, P. and Weiss, R. and Dubourg, V. and Vanderplas, J. and Passos, A. and Cournapeau, D. and Brucher, M. and Perrot, M. and Duchesnay, E.},
  year         = {2011},
  journal      = {Journal of Machine Learning Research},
  volume       = {12},
  pages        = {2825--2830},
  doi          = {10.48550/arXiv.1201.0490}
}

@book{sachdev:qpt:2011,
  title        = {Quantum Phase Transitions},
  author       = {Sachdev, Subir},
  year         = {2011},
  publisher    = {Cambridge University Press},
  edition      = {2nd}
}

@inproceedings{derzkho:2008,
  title        = {Jordan-{{Wigner Fermionization}} and the {{Theory}} of {{Low-Dimensional Quantum Spin Models}}. {{Dynamic Properties}}},
  author       = {Derzkho, O.},
  year         = {2008},
  month        = jun,
  booktitle    = {Condensed Matter Physics in the Prime of the 21{{{\textsuperscript{st}}}} Century},
  publisher    = {World Scientific},
  pages        = {35--87},
  doi          = {10.1142/9789812709455_0002}
}

@article{li:2007,
  title        = {Generating Unexpected Spin Echoes in Dipolar Solids with {{$\pi$}} Pulses},
  author       = {Li, Dale and Dementyev, A. E. and Dong, Yanqun and Ramos, R. G. and Barrett, S. E.},
  year         = {2007},
  month        = may,
  journal      = {Physical Review Letters},
  publisher    = {American Physical Society},
  volume       = {98},
  number       = {19},
  pages        = {190401},
  doi          = {10.1103/PhysRevLett.98.190401}
}

@article{imre:2006,
  title        = {Majority Logic Gate for Magnetic Quantum-Dot Cellular Automata},
  author       = {Imre, A. and Csaba, G. and Ji, L. and Orlov, A. and Bernstein, G. H. and Porod, W.},
  year         = {2006},
  month        = jan,
  journal      = {Science},
  publisher    = {American Association for the Advancement of Science (AAAS)},
  volume       = {311},
  number       = {5758},
  pages        = {205--208},
  doi          = {10.1126/science.1120506},
  issn         = {1095-9203}
}

@article{sutherland:beautiful-models,
  title        = {Beautiful {{Models}}: 70 {{Years}} of {{Exactly Solved Quantum Many-Body Problems}}},
  author       = {Sutherland, Bill and Andrei, N.},
  year         = {2005},
  month        = sep,
  journal      = {Physics Today},
  volume       = {58},
  number       = {9},
  pages        = {58--60},
  doi          = {10.1063/1.2117827},
  issn         = {0031-9228}
}

@article{you:2005,
  title        = {Superconducting {{Circuits}} and {{Quantum Information}}},
  author       = {You, J. Q. and Nori, Franco},
  year         = {2005},
  month        = nov,
  journal      = {Physics Today},
  publisher    = {AIP Publishing},
  volume       = {58},
  number       = {11},
  pages        = {42--47},
  doi          = {10.1063/1.2155757},
  issn         = {1945-0699}
}

@article{floquet:rahav:2003,
  title        = {Effective {{Hamiltonians}} for Periodically Driven Systems},
  author       = {Rahav, Saar and Gilary, Ido and Fishman, Shmuel},
  year         = {2003},
  month        = jul,
  journal      = {Physical Review A: Atomic, Molecular, and Optical Physics},
  publisher    = {American Physical Society},
  volume       = {68},
  number       = {1},
  pages        = {013820},
  doi          = {10.1103/PhysRevA.68.013820}
}

@incollection{tsvelik:2003,
  title        = {Jordan--{{Wigner}} Transformation for Spin {{S}} = 1/2 Models in {{D}} = 1, 2, 3},
  author       = {Tsvelik, Alexei M.},
  year         = {2003},
  booktitle    = {Quantum Field Theory in Condensed Matter Physics},
  publisher    = {Cambridge University Press},
  address      = {Cambridge},
  pages        = {172--178},
  doi          = {10.1017/CBO9780511615832},
  note         = {{ISBN:} 9780511615832}
}

@article{cowburn:2000,
  title        = {Room Temperature Magnetic Quantum Cellular Automata},
  author       = {Cowburn, R. P. and Welland, M. E.},
  year         = {2000},
  month        = feb,
  journal      = {Science},
  publisher    = {American Association for the Advancement of Science (AAAS)},
  volume       = {287},
  number       = {5457},
  pages        = {1466--1468},
  doi          = {10.1126/science.287.5457.1466},
  issn         = {1095-9203}
}

@article{fischler:ransac,
  title        = {Random Sample Consensus: A Paradigm for Model Fitting with Applications to Image Analysis and Automated Cartography},
  author       = {Fischler, Martin A. and Bolles, Robert C.},
  year         = {1981},
  month        = jun,
  journal      = {Communications of The Acm},
  publisher    = {Association for Computing Machinery},
  address      = {New York, NY, USA},
  volume       = {24},
  number       = {6},
  pages        = {381--395},
  doi          = {10.1145/358669.358692},
  issn         = {0001-0782},
  issue_date   = {June 1981}
}

@article{lajzerowicz:pfeuty:1975,
  title        = {Space-time-dependent spin correlation of the one-dimensional {I}sing model with a transverse field},
  author       = {Lajzerowicz, J. and Pfeuty, P.},
  year         = {1975},
  journal      = {Phys. Rev. B},
  volume       = {11},
  pages        = {4560--4562},
  doi          = {10.1103/PhysRevB.11.4560}
}

@article{pfeuty:1970,
  title        = {The One-Dimensional {I}sing Model with a Transverse Field},
  author       = {Pfeuty, Pierre},
  year         = {1970},
  journal      = {Ann. Phys.},
  volume       = {57},
  pages        = {79--90},
  doi          = {10.1016/0003-4916(70)90270-8}
}

@article{yang:1962,
  title        = {Concept of Off-Diagonal Long-Range Order and the Quantum Phases of Liquid He and of Superconductors},
  author       = {Yang, C. N.},
  year         = {1962},
  month        = oct,
  journal      = {Rev. Mod. Phys.},
  publisher    = {American Physical Society},
  volume       = {34},
  pages        = {694--704},
  doi          = {10.1103/RevModPhys.34.694},
  issue        = {4},
  numpages     = {0}
}

@article{lieb:schultz:mattis:1961,
  title        = {Two Soluble Models of an Antiferromagnetic Chain},
  author       = {Lieb, Elliott and Schultz, Theodore and Mattis, Daniel},
  year         = {1961},
  journal      = {Ann. Phys.},
  volume       = {16},
  pages        = {407--466},
  doi          = {10.1016/0003-4916(61)90115-4}
}

@article{bogie:1958,
  title        = {A {{New Method}} in the {{Theory}} of {{Superconductivity}}},
  author       = {Bogoljubov, N. N. and Tolmachov, V. V. and {\v S}irkov, D. V.},
  year         = {1958},
  month        = jan,
  journal      = {Fortschritte der Physik},
  publisher    = {Wiley},
  volume       = {6},
  number       = {11--12},
  pages        = {605--682},
  doi          = {10.1002/prop.19580061102},
  issn         = {0015-8208}
}

@article{valatin:1958,
  title        = {Comments on the Theory of Superconductivity},
  author       = {Valatin, J. G.},
  year         = {1958},
  month        = mar,
  journal      = {Il Nuovo Cimento},
  publisher    = {Springer Science and Business Media LLC},
  volume       = {7},
  number       = {6},
  pages        = {843--857},
  doi          = {10.1007/bf02745589},
  issn         = {1827-6121}
}

@article{penrose:onsager:1956,
  title        = {Bose-Einstein Condensation and Liquid Helium},
  author       = {Penrose, Oliver and Onsager, Lars},
  year         = {1956},
  month        = nov,
  journal      = {Phys. Rev.},
  publisher    = {American Physical Society},
  volume       = {104},
  pages        = {576--584},
  doi          = {10.1103/PhysRev.104.576},
  issue        = {3},
  numpages     = {0}
}

@article{jordan:1928,
  title        = {{\"U}ber Das {{Paulische}} {\"A}quivalenzverbot},
  author       = {Jordan, P. and Wigner, E.},
  year         = {1928},
  month        = sep,
  journal      = {Zeitschrift f{\"u}r Physik},
  publisher    = {Springer Science and Business Media LLC},
  volume       = {47},
  number       = {9--10},
  pages        = {631--651},
  doi          = {10.1007/bf01331938},
  issn         = {1434-601x}
}
\end{document}